\documentclass[reprint,aps,prl,superscriptaddress,nobibnotes]{revtex4-1}%
\usepackage{amsmath}
\usepackage{amssymb}
\usepackage{natbib}
\usepackage{graphicx}
\usepackage{xcolor}
\usepackage{amsfonts}%
\providecommand{\U}[1]{\protect\rule{.1in}{.1in}}

\setcitestyle{numbers,square}

\begin{document}

\title{Chiral Pumping of Spin Waves}

\author{Tao Yu}
\affiliation{Kavli Institute of NanoScience, Delft University of Technology,
	2628 CJ Delft, The Netherlands}
\author{Yaroslav M. Blanter}
\affiliation{Kavli Institute of NanoScience, Delft University of Technology,
	2628 CJ Delft, The Netherlands}
\author{Gerrit E. W. Bauer}
\affiliation{Institute for Materials Research \& WPI-AIMR \& CSRN, Tohoku
	University, Sendai 980-8577, Japan} 
\affiliation{Kavli Institute of
	NanoScience, Delft University of Technology, 2628 CJ Delft, The Netherlands}

\date{\today}

\begin{abstract}
We report a theory for the coherent and incoherent chiral pumping of spin
waves into thin magnetic films through the dipolar coupling with a local
magnetic transducer, such as a nanowire. The ferromagnetic resonance of the
nanowire is broadened by the injection of unidirectional spin waves that
generates a non-equilibrium magnetization in only half of the film. A
temperature gradient between the local magnet and film leads to a
unidirectional flow of incoherent magnons, i.e., a chiral spin Seebeck effect.

\end{abstract}
\maketitle

\textit{Introduction}.---Magnonics and magnon spintronics are fields in which
spin waves---the collective excitations of magnetic order---and their quanta,
magnons, are studied with the purpose of using them as information carriers in
low-power devices \cite{magnonics1,magnonics2,magnonics3,magnonics4}. Magnons
carry angular momentum or \textquotedblleft spin\textquotedblright\ by the
precession direction around the equilibrium state. By angular momentum
conservation the magnon spin couples to electromagnetic waves with only one
polarization \cite{chiral_review}, which can be used to control spin waves
\cite{magnonics1,magnonics2,magnonics3,magnonics4}. Surface spin waves or
Damon-Eshbach (DE) modes have also a handedness or chirality, i.e. their
linear momentum is fixed by the outer product of surface normal and
magnetization direction \cite{Walker,DE,spin_waves_book,new_book}. Alas,
surface magnons have small group velocity, are dephased easily by surface
roughness \cite{surface_roughness}, and exist only in sufficiently thick
magnetic films, which explains why they have not been employed for
applications in magnonic devices \cite{logic}\textit{.}

The favored material for magnonics is the ferrimagnetic insulator yttrium iron
garnet (YIG) with record low magnetization damping and high Curie temperature
\cite{low_damping_nanometer}. Spin waves in YIG films can be classified by the
interaction that governs their dispersion as a function of wave vector to be
of the dipolar, dipolar-exchange and exchange type with energies from a few
GHz to many THz \cite{magnonics1,magnonics2,magnonics3,magnonics4}.
Long-wavelength dipolar spin waves can be coherently excited by microwaves and
travel over centimeters \cite{centimeter}, but suffer from low group
velocities. Exchange spin waves have much higher group velocity, but they can
often be excited only incoherently by thermal or electric actuation via
metallic contacts \cite{Ludo}. They are also scattered easily, leading to
diffuse transport with reduced (micrometer) length scale. The dipolar-exchange
spin waves are potentially most suitable for coherent information technologies
by combining speed with long lifetime. Recently, short-wavelength
dipolar-exchange spin waves have been coherently excited in magnetic films by
attaching transducers in the form of thin and narrow ferromagnetic wires or
wire arrays with high resonance frequencies
\cite{chiral_simulation,Dirk_2013,CoFeB_YIG,Co_YIG,Haiming_NC,Haiming_PRL}.
The dipolar interaction dominantly couples the transducer dynamics with the
film, but in direct contact interface exchange and spin transfer torque may
also play a role. Micromagnetic simulations \cite{chiral_simulation} revealed
that the AC dipolar field emitted by a magnetic wire antenna can excite spin
waves in a magnetic film with magnetization normal to the wire, but no
physical arguments or experiments supported this finding. Recently, almost
perfectly chiral excitation of exchange spin waves was observed in thin YIG
films with Co or Ni nanowire gratings with collinear magnetizations
\cite{Yu2,Yu1}.

\begin{figure}[th]
{\includegraphics[width=9.1cm]{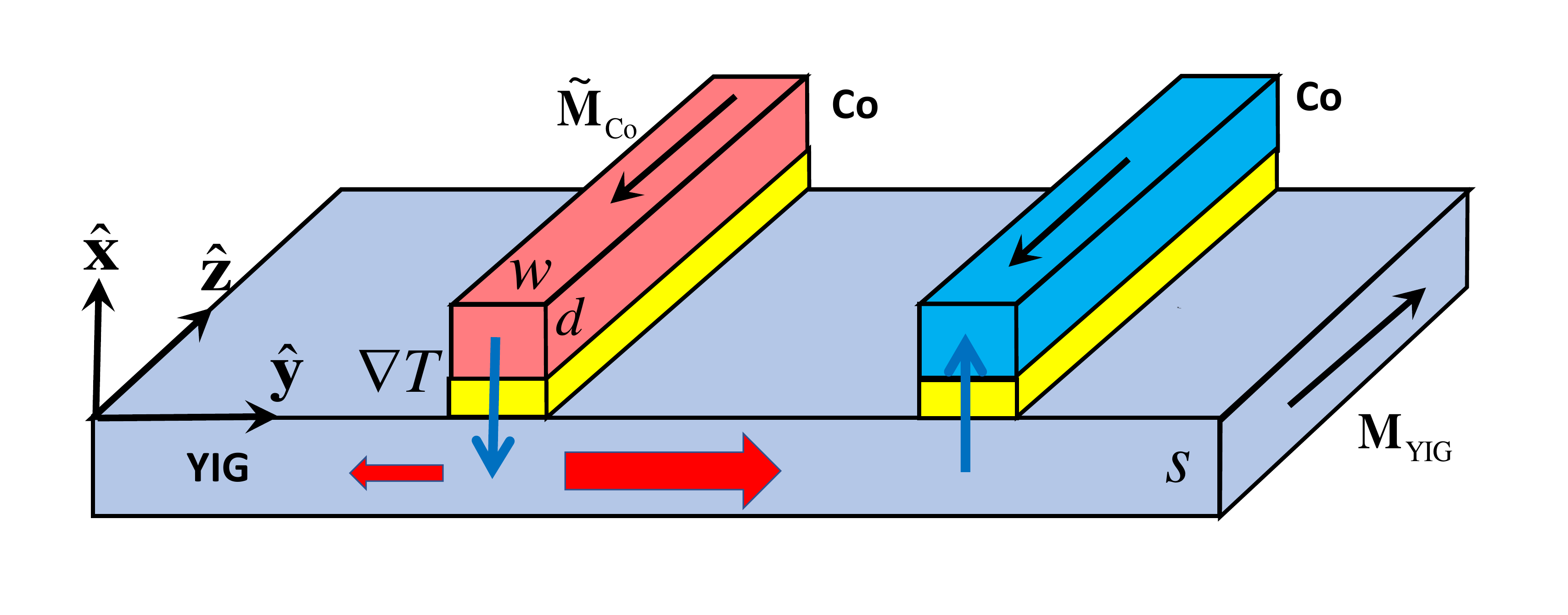}}\caption{Chiral Spin
Seebeck effect. A thin non-magnetic spacer between the YIG film and Co
nanowire (optionally) suppresses the exchange interaction. The effect is
maximal for the antiparallel magnetization (see text). The magnitude of the
magnon currents pumped into the $\pm\mathbf{\hat{y}}$ directions is indicated
by the size of the red arrows. Another Co nanowire (the blue one) is suggested
to detect the population or temperature of magnon with long-wavelength.}%
\label{incoherent2}%
\end{figure}

The chiral excitation of spin waves \cite{Yu1} corresponds to a robust and
switchable exchange magnon current generated by microwaves. The generation of
DC currents by AC forces in the absence of a DC bias is referred to as
\textquotedblleft pumping\textquotedblright\ \cite{pumping_Buttiker}. Spin
pumping\ is the injection of a spin current by the magnetization dynamics of a
magnet into a contact normal metal by the interface exchange interaction
\cite{spin_pumping_electron,non_local}. We therefore call generation of
unidirectional spin waves by the dynamics of a proximity magnetic wire
\textit{chiral spin pumping}. Here we present a semi-analytic theory of chiral
spin pumping for arbitrary magnetic configurations. We distinguish coherent
pumping by applied microwaves from the incoherent (thermal) pumping by a
temperature difference, i.e. the chiral spin Seebeck effect
\cite{Spin_seebeck_exp,Spin_seebeck_theory1,Spin_seebeck_theory2,Spin_caloritronics}
as shown schematically in Fig.~\ref{incoherent2}. The former has been studied
by microwave transmission spectroscopy \cite{Yu2,Yu1}. Both effects can be
observed also electrically via the inverse spin Hall effect in heavy metal
contacts, but we focus here on the more efficient inductive detection scheme.

Chiral spin pumping turns out to be very anisotropic. When spin waves
propagate perpendicular to the magnetization with opposite momenta, their
dipolar fields vanish on opposite sides of the film; when propagating parallel
to the magnetization, their dipolar field is chiral, i.e.,
polarization-momentum locked. Purely chiral coupling between magnons can be
achieved in the former case without constraints on the degree of polarization
of the local magnet. We also find that the pumping by dipolar interaction is
chiral in both momentum and real space, i.e., in the configuration of
Fig.~\ref{incoherent2} \textit{unidirectional} spin waves are excited in
\textit{half} of the film.

\textit{Origin of the chiral coupling}.---The dynamic dipolar coupling of
magnetization $\mathbf{\tilde{M}}$ of the local magnet with that of a film
$\mathbf{M}$ by the Zeeman interaction with the respective dipolar magnetic
fields $\mathbf{h}$ and $\mathbf{\tilde{h}}$ \cite{Landau}
\begin{equation}
\hat{H}_{\mathrm{int}}/\mu_{0}=-\int\mathbf{\tilde{M}}(\mathbf{r}%
,t)\cdot\mathbf{h}(\mathbf{r},t)d\mathbf{r}=-\int\mathbf{M}(\mathbf{r}%
,t)\cdot\mathbf{\tilde{h}}(\mathbf{r},t)d\mathbf{r}, \label{Hint}%
\end{equation}
where $\mu_{0}$ is the vacuum permeability. We focus here on circularly
polarized exchange spin waves in a magnetic film with thickness $s$ at
frequency $\omega$ and in-plane wave vector $\mathbf{k}=k_{y}\hat{\mathbf{y}%
}+k_{z}\hat{\mathbf{z}}$ in the coordinate system defined in
Fig.~\ref{incoherent2} (the general case is treated in the Supplemental
Material (SM) Sec.~I.A \cite{supplement}). Classically, $M_{x}(\mathbf{r}%
,t)=m_{R}^{\mathbf{k}}(x)\cos(\mathbf{k}\cdot\pmb{\rho}-\omega t)$ and
$M_{y}(\mathbf{r},t)\equiv-m_{R}^{\mathbf{k}}(x)\sin(\mathbf{k}\cdot
\pmb{\rho}-\omega t)$ describe the precession around the equilibrium
magnetization modulated in the $\hat{\mathbf{z}}$-direction, where
$m_{R}^{\mathbf{k}}(x)$ is the time-independent amplitude normal to the film
and $\pmb{\rho}=y\hat{\mathbf{y}}+z\hat{\mathbf{z}}.$ The dipolar field
outside the film generated by the spin waves
\begin{equation}
h_{\beta}(\mathbf{r},t)=\frac{1}{4\pi}\partial_{\beta}\partial_{\alpha}\int
d\mathbf{r}^{\prime}\frac{M_{\alpha}(\mathbf{r}^{\prime},t)}{|\mathbf{r}%
-\mathbf{r}^{\prime}|}, \label{dipolar_field_definition}%
\end{equation}
in the summation convention over repeated Cartesian indices $\alpha
,\beta=\{x,y,z\}$ \cite{Landau}, becomes
\begin{align}
\left(
\begin{array}
[c]{c}%
h_{x}(\mathbf{r},t)\\
h_{y}(\mathbf{r},t)\\
h_{z}(\mathbf{r},t)
\end{array}
\right)   &  =\left(
\begin{array}
[c]{c}%
\left(  k+\eta k_{y}\right)  \cos\left(  \mathbf{k}\cdot\pmb{\rho}-\omega
t\right) \\
\left(  \frac{k_{y}^{2}}{k}+\eta k_{y}\right)  \sin\left(  \mathbf{k}%
\cdot\pmb{\rho}-\omega t\right) \\
k_{z}\left(  \frac{k_{y}}{k}+\eta\right)  \sin\left(  \mathbf{k}%
\cdot\pmb{\rho}-\omega t\right)
\end{array}
\right) \nonumber\\
&  \times\frac{1}{2}e^{-\eta kx}\int dx^{\prime}m_{R}^{\mathbf{k}}\left(
x^{\prime}\right)  e^{\eta kx^{\prime}}, \label{dipolar_field}%
\end{align}
where the spatial integral is over the film thickness $s$. $x>0$ ($x<-s$) is
the case with the dipolar field above (below) the film and $\eta=1$ ($-1$)
when $x>0$ ($x<-s$), $k=\left\vert \mathbf{k}\right\vert $. The interaction
Hamiltonian (\ref{Hint}) for a wire with thickness $d$ and width $w$
\cite{Landau} reduced to%
\begin{equation}
\hat{H}_{\mathrm{int}}(t)=-\mu_{0}\int_{0}^{d}\hat{\tilde{M}}_{\beta
}(x,\pmb{\rho},t)\hat{h}_{\beta}(x,\pmb{\rho},t)dxd\pmb{\bf \rho}.
\label{interaction}%
\end{equation}

\begin{figure}[th]
{\includegraphics[width=7.9cm]{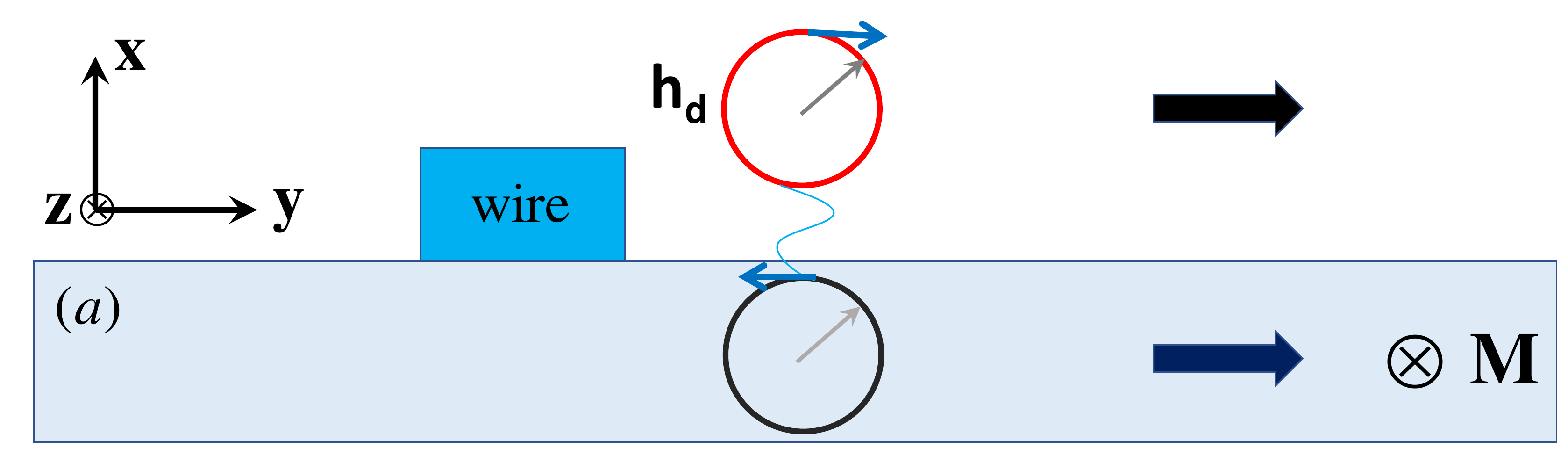}}
{\includegraphics[width=7.9cm]{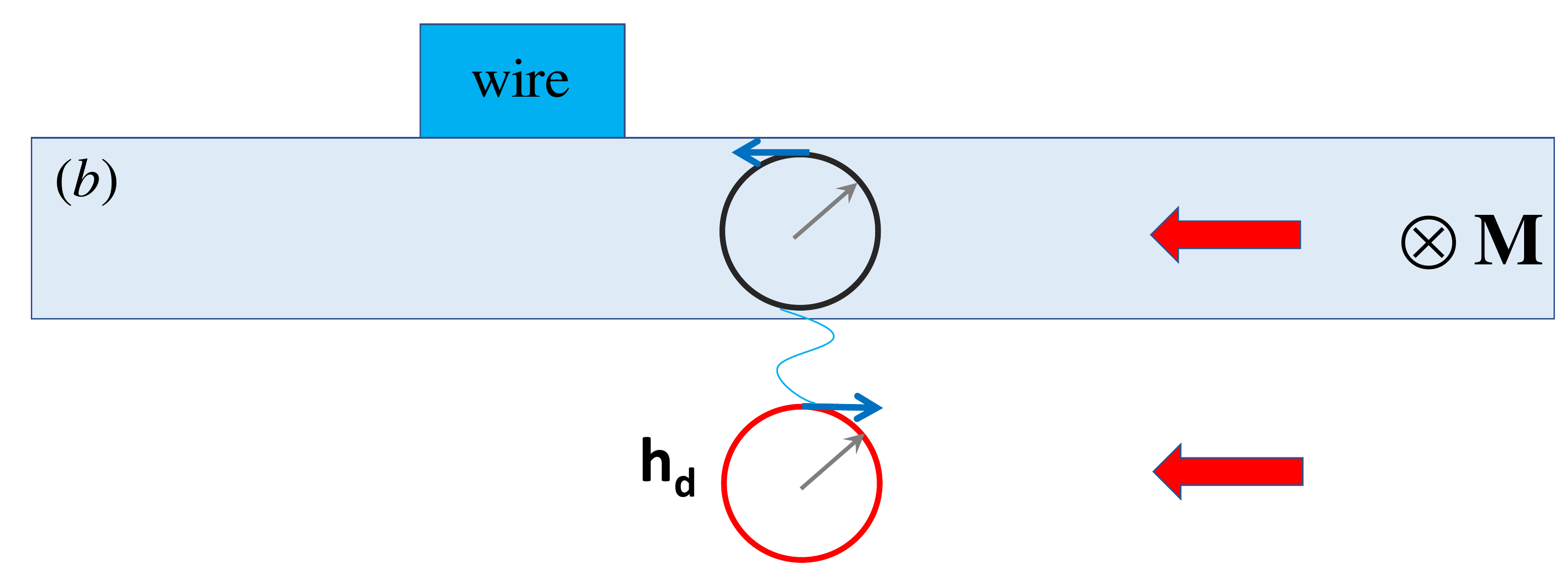}}\caption{Half-space dipolar fields
generated by spin waves propagating normal to the (equilibrium) magnetization
of an in-plane magnetized film ($\mathbf{M}_{s}\parallel\hat{\mathbf{z}}$).
The fat black arrow in (a) and red arrows in (b) indicate the spin wave
propagation direction. The black (red) circles are the precession cones of the
film magnetization (corresponding dipolar field) and precession direction is
indicated by thin blue arrows.}%
\label{fig:dipolar_field}%
\end{figure}

The spin waves in the film with $k_{z}=0$ propagate normal to the wire with
dipolar field $h_{z}=0$. The distribution of the dipolar field above and below
the film then strongly depends on the wave vector direction: the dipolar field
generated by the right (left) moving spin waves vanishes below (above) the
film \cite{Yu1} and precesses in the opposite direction of the magnetization
as sketched in Fig.~\ref{fig:dipolar_field}. The magnetization in the wire
precesses in a direction governed by the magnetization direction and couples
only to spin waves with finite dipolar field amplitude in the wire and matched
precession \cite{Yu1,Yu2}. We thus understand without calculations that the
dipolar coupling is chiral and the time-averaged coupling strength is
maximized when the magnetizations of the film and wire are
antiparallel.\textit{\ }

Spin waves in the film propagating \textit{parallel} to the magnetization
($k_{y}=0\rightarrow h_{y}=0$) may also couple chirally to the local magnet,
but by a different mechanism. According to Eq.~(\ref{dipolar_field}),
$h_{x}\propto\left\vert k_{z}\right\vert \cos\left(  k_{z}z-\omega t\right)  $
and $h_{z}\propto\eta k_{z}\sin\left(  k_{z}z-\omega t\right)  $. Above the
film, the dipolar fields with positive (negative) $k_{z}$ are left (right)
circularly polarized, respectively, while below the film, the polarizations
are reversed. These spin waves couple with the magnet on one side of the film
only when its transverse magnetization dynamics is right or left circularly
polarized \cite{chiral_simulation}.

A circularly polarized uniform precession in the nanowire always couples
chirally with the spin waves in the film (see SM Secs.~I.B and II
\cite{supplement}) for all angles between magnetizations in film and nanowire
irrespective of their polarization. When the nanowire Kittel mode is
elliptical, the directionality vanishes for one specific angle $\theta_{c}$.
When the nanowire Kittel mode is fully circularly polarized, the coupling
strength vanishes and the critical angle $\theta_{c}=0$. With $w>d$ and weak
magnetic field bias, $\theta_{c}\simeq\arccos\left(  \sqrt{d/w}\right)  $ (see
SM Sec.~II \cite{supplement}) and the chirality can be controlled by weak
in-plane magnetic fields.

\textit{General formalism}.---Here we formulate the general problem of the
dynamic dipolar coupling between a nanowire with equilibrium magnetization at
an angle $\theta$ that is in contact with an extended thin magnetic film. At
resonance, microwaves populate preferentially the collective
(\textquotedblleft Kittel\textquotedblright) modes \cite{Kittel_1948}, while a
finite temperature populates all magnon modes with a Planck distribution. We
focus here on the collinear (parallel and antiparallel) configurations,
deferring the derivations and discussions of general situations to the SM
Sec.~II \cite{supplement}. For convenience, we formulate the problem in second quantization.

For sufficiently small amplitudes, the Cartesian components $\beta\in\left\{
x,y\right\}  $ of the magnetization dynamics of film ($\mathbf{\hat{M}}$) and
nanowire ($\mathbf{\hat{\tilde{M}}}$) can be expanded into magnon creation and
annihilation operators \cite{Kittel_book,HP,surface_roughness},
\begin{align}
\hat{M}_{\beta}(\mathbf{r})  &  =-\sqrt{2M_{s}\gamma\hbar}\sum_{\mathbf{k}%
}\left[  m_{\beta}^{\left(  \mathbf{k}\right)  }(x)e^{i{\mathbf{k}}%
\cdot\pmb{\rho}}\hat{\alpha}_{\mathbf{k}}+\mathrm{h.c.}\right]  ,\nonumber\\
\hat{\tilde{M}}_{\beta}(\mathbf{r})  &  =-\sqrt{2\tilde{M}_{s}\gamma\hbar}%
\sum_{k_{z}}\left[  \tilde{m}_{\beta}^{(k_{z})}(x,y)e^{ik_{z}z}\hat{\beta
}_{k_{z}}+\mathrm{h.c.}\right]  , \label{expansion}%
\end{align}
where $M_{s}$ and $\tilde{M}_{s}$ are the saturation magnetizations, $-\gamma$
is the gyromagnetic ratio, $m_{\beta}^{\left(  \mathbf{k}\right)  }(x)$ and
$\tilde{m}_{\beta}^{(k_{z})}(x,y)$ are the spin wave amplitudes across the
film and nanowire, and $\hat{\alpha}_{\mathbf{k}}$ and $\hat{\beta}_{k_{z}}$
denote the magnon (annihilation) operator in the film and nanowire, respectively.

We are mainly interested in high-quality ultrathin films and nanowires with
$s,d\gtrsim\mathcal{O}\left(  10~\mathrm{nm}\right)  $ and nanowire width
$w\gtrsim\mathcal{O}\left(  50~\mathrm{nm}\right)  $, such that the
magnetization across the film and nanowire (centered at $y_{0}\hat{\mathbf{y}%
}$) are nearly homogeneous: $m_{\beta}^{\left(  \mathbf{k}\right)  }(x)\approx
m_{\beta}^{\left(  \mathbf{k}\right)  }\Theta(-x)\Theta(x+s)$ and $\tilde
{m}_{\beta}^{\left(  k_{z}\right)  }(x,y)\approx\tilde{m}_{\beta}^{\left(
k_{z}\right)  }\Theta(x)\Theta(-x+d)\Theta(y-y_{0}+w/2)\Theta(-y+y_{0}+w/2)$,
with $\Theta(x)$ the Heaviside step function \cite{Yu1,Yu2,Haiming_PRL}.
Disregarding higher magnon subbands turns out to be a good approximation even
at higher temperatures because of the strong mode selectivity of the dipolar
coupling \cite{supplement}. Here we disregard interface exchange, which
appears to play only a minor role \cite{Yu1,Yu2}. The system Hamiltonian then
becomes
\begin{align}
\hat{H}/\hbar &  =\sum_{\mathbf{k}}\omega_{\mathbf{k}}\hat{\alpha}%
_{\mathbf{k}}^{\dagger}\hat{\alpha}_{\mathbf{k}}+\sum_{k_{z}}\tilde{\omega
}_{k_{z}}\hat{\beta}_{k_{z}}^{\dagger}\hat{\beta}_{k_{z}}\nonumber\\
&  +\sum_{\mathbf{k}}\left(  g_{\mathbf{k}}e^{-ik_{y}y_{0}}\hat{\alpha
}_{\mathbf{k}}^{\dagger}\hat{\beta}_{k_{z}}+g_{\mathbf{k}}^{\ast}%
e^{ik_{y}y_{0}}\hat{\beta}_{k_{z}}^{\dagger}\hat{\alpha}_{\mathbf{k}}\right)
,
\end{align}
where $\omega_{\mathbf{k}}$ and $\tilde{\omega}_{k_{z}}$ are the frequencies
of spin waves in the film and nanowire and, with
Eqs.~(\ref{dipolar_field_definition}), (\ref{interaction}) and
(\ref{expansion}), the coupling
\begin{equation}
g_{\mathbf{k}}=F(\mathbf{k})\left(  m_{x}^{(\mathbf{k})\ast},m_{y}%
^{(\mathbf{k})\ast}\right)  \left(
\begin{array}
[c]{cc}%
|\mathbf{k}| & ik_{y}\\
ik_{y} & -k_{y}^{2}/|\mathbf{k}|
\end{array}
\right)  \left(
\begin{array}
[c]{c}%
\tilde{m}_{x}^{(k_{z})}\\
\tilde{m}_{y}^{(k_{z})}%
\end{array}
\right)  , \label{coupling_constant}%
\end{equation}
with $F(\mathbf{k})=-\mu_{0}\gamma\sqrt{M_{s}\tilde{M}_{s}/L}\phi\left(
\mathbf{k}\right)  $. The form factor $\phi\left(  \mathbf{k}\right)
=2\sin(k_{y}w/2)(1-e^{-kd})(1-e^{-ks})/(k_{y}{k^{2}})$ couples spin waves with
wavelengths of the order of the nanowire width (mode selection) and
$\lim_{\mathbf{\ k}\rightarrow0}\phi\left(  \mathbf{k}\right)  =wsd$. Exchange
waves are right-circularly polarized with $m_{y}^{(k_{y})}=im_{x}^{(k_{y})}$
and the coupling is perfectly chiral $g_{-|k_{y}|}=0$ (refer to
Fig.~\ref{fig:dipolar_field}).

The linear response to microwave and thermal excitations can be described by
the input-output theory \cite{input_output1,input_output2} and by a Kubo
formula (see SM Sec.~III \cite{supplement}). Let $\hat{p}_{k_{z}}(t)=\int%
\hat{p}_{k_{z}}(\omega)e^{-i\omega t}d\omega/(2\pi)$ be a microwave photon
input with magnetic field $\propto(\hat{p}_{k_{z}}(t)+\hat{p}_{k_{z}}%
^{\dagger}(t))$ centered at the frequency $\tilde{\omega}_{k_{z}}:$%
\textit{\ }$\langle\hat{p}_{k_{z}}(\omega)\rangle\rightarrow2\pi
\mathit{\mathcal{D}}\delta(\omega-\tilde{\omega}_{k_{z}})$ with amplitude
$\mathcal{D}$, the equations of motion are \cite{input_output1,input_output2}%
\
\begin{align}
\frac{d\hat{\beta}_{k_{z}}}{dt}  &  =-i\tilde{\omega}_{k_{z}}\hat{\beta
}_{k_{z}}(t)-\sum_{k_{y}}ig_{\mathbf{k}}^{\ast}e^{ik_{y}y_{0}}\hat{\alpha
}_{\mathbf{k}}(t)\nonumber\\
&  -\left(  \frac{\tilde{\kappa}_{k_{z}}}{2}+\frac{\zeta_{k_{z}}}{2}\right)
\hat{\beta}_{k_{z}}(t)-\sqrt{\tilde{\kappa}_{k_{z}}}\hat{\tilde{N}}_{k_{z}%
}(t)-\sqrt{\zeta_{k_{z}}}\hat{p}_{k_{z}}(t),\\
\frac{d\hat{\alpha}_{\mathbf{k}}}{dt}  &  =-i\omega_{\mathbf{k}}\hat{\alpha
}_{\mathbf{k}}(t)-ig_{\mathbf{k}}e^{-ik_{y}y_{0}}\hat{\beta}_{k_{z}}%
(t)-\frac{\kappa_{\mathbf{k}}}{2}\hat{\alpha}_{\mathbf{k}}(t)\nonumber\\
&  -\sqrt{\kappa_{\mathbf{k}}}\hat{N}_{\mathbf{k}}(t),
\end{align}
where $\tilde{\kappa}_{k_{z}}\equiv2\tilde{\chi}\tilde{\omega}_{k_{z}}%
\ $($\kappa_{\mathbf{k}}\equiv2\chi\omega_{\mathbf{k}}$) is the damping rates
in terms of the Gilbert damping constant $\tilde{\chi}\ (\chi)$ in the
nanowire (film) and $\zeta_{k_{z}}$ is the radiative damping\textit{.} The
thermal environment of the magnetic film causes fluctuations $\hat
{N}_{\mathbf{k}}$ \cite{input_output2} generated by a Markovian process that
obeys the (quantum) fluctuation-dissipation theorem with $\langle{\hat
{N}_{\mathbf{k}}}\rangle=0$ and $\langle{\hat{N}_{\mathbf{k}}^{\dagger}%
(t)\hat{N}_{\mathbf{k}^{\prime}}(t^{\prime})}\rangle=n_{\mathbf{k}}%
\delta(t-t^{\prime})\delta_{\mathbf{k}\mathbf{k}^{\prime}}$. $n_{\mathbf{k}%
}=1/\left\{  \exp\left[  {\hbar\omega_{\mathbf{k}}}/({k_{B}T_{2}})\right]
-1\right\}  $ is the magnon population at temperature $T_{2}$ of film, and
$k_{B}T_{2}$ should be larger than $\hbar\omega_{\mathbf{k}}$. In frequency
space with $\hat{A}(t)=\int d\omega/({2\pi})\hat{A}(\omega)e^{-i\omega t}$,
$\left\langle \hat{N}_{\mathbf{k}}^{\dagger}(\omega)\hat{N}_{\mathbf{k}%
^{\prime}}(\omega^{\prime})\right\rangle =2\pi\delta(\omega-\omega^{\prime
})n_{\mathbf{k}}\delta_{\mathbf{k}\mathbf{k}^{\prime}}$. The thermal
fluctuations $\hat{\tilde{N}}_{k_{z}}$ in the nanowire are characterized by a
different temperature $T_{1}$ and thermal magnon distribution $\tilde
{n}_{k_{z}}=1/\left\{  \exp\left[  {\hbar\tilde{\omega}_{k_{z}}}/({k_{B}T_{1}%
})\right]  -1\right\}  $. The solutions \textit{ }
\begin{align}
\hat{\beta}_{k_{z}}(\omega)  &  =\frac{i\sum\limits_{k_{y}}\gamma_{\mathbf{k}%
}G_{\mathbf{k}}\hat{N}_{\mathbf{k}}(\omega)-\sqrt{\tilde{\kappa}_{k_{z}}}%
\hat{\tilde{N}}_{k_{z}}(\omega)-\sqrt{\zeta_{k_{z}}}\hat{p}_{k_{z}}(\omega
)}{-i(\omega-\tilde{\omega}_{k_{z}})+\frac{\tilde{\kappa}_{k_{z}}}{2}%
+\frac{\zeta_{k_{z}}}{2}+i\sum_{k_{y}}|g_{\mathbf{k}}|^{2}G_{\mathbf{k}%
}\left(  \omega\right)  },\nonumber\\
\hat{\alpha}_{\mathbf{k}}(\omega)  &  =G_{\mathbf{k}}\left(  \omega\right)
\left(  g_{\mathbf{k}}e^{-ik_{y}y_{0}}\hat{\beta}_{k_{z}}(\omega
)-i\sqrt{\kappa_{\mathbf{k}}}\hat{N}_{\mathbf{k}}(\omega)\right)  ,
\label{Kittel_excitation}%
\end{align}
with Green function $G_{\mathbf{k}}\left(  \omega\right)  =\left(
(\omega-\omega_{\mathbf{k}})+i\kappa_{\mathbf{k}}/2\right)  ^{-1}$ and
$\gamma_{\mathbf{k}}=ig_{\mathbf{k}}^{\ast}e^{ik_{y}y_{0}}\sqrt{\kappa
_{\mathbf{k}}}$, reveal that the thermal fluctuations in both wire
($\hat{\tilde{N}}_{k_{z}}$) and film ($\hat{N}_{\mathbf{k}}$) affect
$\hat{\beta}_{k_{z}}(\omega)$. Moreover, the interaction enhances the damping
of nanowire spin waves by $\delta\tilde{\kappa}_{k_{z}}=2\pi\sum_{k_{y}%
}|g_{\mathbf{k}}|^{2}\delta(\tilde{\omega}_{k_{z}}-\omega_{\mathbf{k}})$ to
$\tilde{\kappa}_{k_{z}}^{\prime}$, and shifts the frequency to $\tilde{\omega
}_{k_{z}}^{\prime}$. Chiral pumping can be realized by coherent microwave
excitation or the incoherent excitation by a temperature difference between
the local magnet and film, as shown in the following.

\textit{Coherent chiral pumping}.---A uniform microwave field excites only the
Kittel mode ($k_{z}=0$) in the nanowire but not the film. Spin waves in the
film with finite $k_{y}\equiv q$ are excited indirectly by the inhomogeneous
stray field of the wire\textit{.} The coherent chiral pumping by microwaves at
thermal equilibrium with $T_{1}=T_{2}\equiv T_{0}$ in the time and wave number
domain reads
\begin{equation}
\hat{\alpha}_{q}(t)=\int\frac{d\omega}{2\pi}\frac{e^{-i\omega t}%
ig_{q}e^{-iqy_{0}}}{-i(\omega-\omega_{q})+\frac{\kappa_{q}}{2}}\frac
{\sqrt{\zeta_{0}}\hat{p}_{0}(\omega)}{-i(\omega-\tilde{\omega}_{0}^{\prime
})+\frac{\tilde{\kappa}_{0}^{\prime}}{2}+\frac{\zeta_{0}}{2}}.
\end{equation}
Since the magnons are coherently excited their number is $\langle\hat{\alpha
}_{q}^{\dagger}(t)\hat{\alpha}_{q}(t)\rangle=\langle\hat{\alpha}_{q}^{\dagger
}(t)\rangle\langle\hat{\alpha}_{q}(t)\rangle$. In the absence of damping,
$\kappa_{q}$ is a positive infinitesimal that safeguards causality. A resonant
input $\langle\hat{p}_{0}\rangle=2\pi\mathcal{D}\delta(\omega-\tilde{\omega
}_{0})$ excites a film magnetization in position space
\begin{align}
\delta{M}_{\beta}(\mathbf{r},t)  &  =\sqrt{2M_{s}\gamma\hbar}\mathcal{D}%
\frac{e^{-i\tilde{\omega}_{0}t}\sqrt{\zeta_{0}}}{-i(\tilde{\omega}_{0}%
-\tilde{\omega}_{0}^{\prime})+(\tilde{\kappa}_{0}^{\prime}+\zeta_{0}%
)/2}\nonumber\\
&  \times\sum_{q}m_{\beta}^{\left(  q\right)  }(x)\frac{g_{q}e^{iq(y-y_{0})}%
}{(\tilde{\omega}_{0}-\omega_{q})+i\kappa_{q}/2}+\mathrm{h.c.}.
\end{align}
The denominator $\tilde{\omega}_{0}-\omega_{q}+i\kappa_{q}/2\ $vanishes for
$q_{\pm}=\pm(q_{\ast}+i\delta_{\Gamma})$ in the complex plane with $q_{\ast
}>0$ and inverse propagation length $\delta_{\Gamma}$. Closing the contour, we
obtain
\begin{align}
\delta{M}_{\beta}(\mathbf{r})  &  =\sqrt{2M_{s}\gamma\hbar}\mathcal{D}\frac
{1}{v_{q_{\ast}}}\frac{e^{-i\tilde{\omega}_{0}t}\sqrt{\zeta_{0}}}%
{(\tilde{\omega}_{0}-\tilde{\omega}_{0}^{\prime})+i(\tilde{\kappa}_{0}%
^{\prime}+\zeta_{0})/2}\nonumber\\
&  \times\left\{
\begin{array}
[c]{c}%
m_{\beta}^{(q_{\ast})}(x)g_{q_{\ast}}e^{iq_{+}(y-y_{0})}+\mathrm{h.c.}\\
m_{\beta}^{(-q_{\ast})}(x)g_{-q_{\ast}}e^{iq_{-}(y-y_{0})}+\mathrm{h.c.}%
\end{array}
\text{ for }%
\begin{array}
[c]{c}%
y>y_{0}\\
y<y_{0}%
\end{array}
\right.  , \label{excited_M}%
\end{align}
where $v_{q_{\ast}}=\left.  \partial\omega_{q}/\partial q\right\vert
_{q_{\ast}}$ is the magnon group velocity. For perfect chiral coupling
$g_{-q_{\ast}}=0$ only the magnetization in half space $y>y_{0}$ can be
excited, which implies handedness also in position space.

The coherent chiral pumping can be directly observed by microwave transmission
spectra \cite{Dirk_2013,Haiming_NC,Haiming_PRL}. We let here two nanowires at
$\mathbf{r}_{1}=R_{1}\hat{\mathbf{y}}$ and $\mathbf{r}_{2}=R_{2}%
\hat{\mathbf{y}}$ act as excitation and detection transducers. The spin wave
transmission amplitude as derived and calculated in the SM Sec.~IV
\cite{supplement} reads
\begin{equation}
S_{21}(\omega)=\frac{\left[  1-S_{11}(\omega)\right]  \sum_{q}iG_{q}\left(
\omega\right)  |g_{q}|^{2}e^{iq(R_{2}-R_{1})}}{-i(\omega-\tilde{\omega}%
_{0})+\tilde{\kappa}_{0}/2+i\sum_{q}G_{q}\left(  \omega\right)  |g_{q}|^{2}},
\end{equation}
and the reflection amplitude $S_{11}(\omega)$ is given in the SM. Chirality
enters via the phase factor $e^{iq(R_{2}-R_{1})}$: When $g_{-|q|}=0$, spin and
microwaves are transmitted from 1 to 2 only when $R_{2}>R_{1}$.

\textit{Incoherent chiral pumping}.---Spin waves can be incoherently excited
by locally heating the nanowire, e.g. by the Joule heating due to an applied
current \cite{Ludo}. In the absence of microwaves $\hat{p}_{k_{z}}=0$, the
magnon distribution of the film reads%
\begin{align}
f(\mathbf{k})  &  \equiv\langle\hat{\alpha}_{\mathbf{k}}^{\dagger}%
(t)\hat{\alpha}_{\mathbf{k}}(t)\rangle\nonumber\\
&  =n_{\mathbf{k}}+\frac{|g_{\mathbf{k}}|^{2}}{(\omega_{\mathbf{k}}%
-\tilde{\omega}_{k_{z}}^{\prime})^{2}+(\tilde{\kappa}_{k_{z}}^{\prime}/2)^{2}%
}\frac{\tilde{\kappa}_{k_{z}}}{\kappa_{\mathbf{k}}}\left(  \tilde{n}_{k_{z}%
}-n_{\mathbf{k}}\right)  .
\end{align}
When $T_{1}>T_{2}$, magnons are injected from the local magnet into the film.
When the coupling is chiral with $g_{\mathbf{k}}\neq g_{-\mathbf{k}}$, the
distribution of magnons is asymmetric, $f(\mathbf{k})\neq f(-\mathbf{k})$,
i.e. carries a unidirectional spin current $I\propto\sum_{\mathbf{k}%
}v_{\mathbf{k}}f(\mathbf{k})$, which in turn generates a magnon accumulation
in the detector magnet.

All occupied modes in the local magnet contribute to the excitation of the
film. In position space
\begin{equation}
\hat{\alpha}(\pmb{\rho},t)=\int\frac{d\omega}{2\pi}e^{-i\omega t}%
\sum_{\mathbf{k}}e^{i\mathbf{k}\cdot\pmb{\rho}}\hat{\alpha}_{\mathbf{k}%
}(\omega),
\end{equation}
and the excited magnon density for $y>y_{0}$ in a high-quality film with
$\kappa_{\mathbf{k}}\rightarrow0_{+}$ reads
\begin{align}
&  \delta\rho_{>}\equiv\left\langle \hat{\alpha}^{\dagger}(\pmb{\rho},t)\hat
{\alpha}(\pmb{\rho},t)\right\rangle |_{y>y_{0}}-\sum_{\mathbf{k}}%
n_{\mathbf{k}}\nonumber\\
&  =\sum_{k_{z}}\int\frac{d\omega}{2\pi}(\tilde{n}_{k_{z}}-n_{\mathbf{k}%
_{\omega}})\frac{|g_{\mathbf{k}_{\omega}}|^{2}}{v_{\mathbf{k}_{\omega}}^{2}%
}\frac{\tilde{\kappa}_{k_{z}}}{(\omega-\tilde{\omega}_{k_{z}}^{\prime}%
)^{2}+(\tilde{\kappa}_{k_{z}}^{\prime}/2)^{2}},
\end{align}
where $\mathbf{k}_{\omega}=q_{\omega}\hat{\mathbf{y}}+k_{z}\hat{\mathbf{z}}$
and $q_{\omega}$ is the positive root of $\omega_{q_{\omega},k_{z}}=\omega$.
For weak magnetic damping in the wire $\tilde{\kappa}_{m}\ll\tilde{\omega}%
_{m}$, the r.h.s reduces to $\delta\rho_{>}\approx\sum_{k_{z}}\left(
{|g_{q_{\ast},k_{z}}|^{2}}/{v_{q_{\ast},k_{z}}^{2}}\right)  (\tilde{n}_{k_{z}%
}-n_{q_{\ast},k_{z}})$ and $\omega_{q_{\ast},k_{z}}=\tilde{\omega}_{k_{z}}$.
For $y<y_{0}$, $\delta\rho_{<}\approx\sum_{k_{z}}\left(  {|g_{-q_{\ast},k_{z}%
}|^{2}}/{v_{q_{\ast},k_{z}}^{2}}\right)  (\tilde{n}_{k_{z}}-n_{q_{\ast},k_{z}%
})\neq\delta\rho_{>}$. We conclude that the thermal injection via chiral
coupling leads to different magnon densities on both sides of the nanowire.
This is a chiral equivalent of the conventional spin Seebeck effect
\cite{Spin_seebeck_exp,Spin_seebeck_theory1,Spin_seebeck_theory2,Spin_caloritronics}%
.

The chiral pumping of magnons can be detected inductively via microwave
emission of a second magnetic wire, by Brillouin light scattering
\cite{Slavin_BLS,Yu2}, NV center magnetometry \cite{NV_center}, and
electrically by the inverse spin Hall effect \cite{Ludo}. The incoherent
excitation couples strongly only with the long wavelength modes that
propagate ballistically over large distances, and the effect is most efficiently
detected by a mode-selective spectroscopy. The thermally excited population of
the Kittel mode in the (right) detection transducer reads (see derivation and
discussion in SM Sec.~V \cite{supplement})
\begin{equation}
\delta\rho_{R}=\int\frac{d\omega}{2\pi}\frac{\Gamma_{1}^{2}\tilde{\kappa}%
_{0}(n_{L}-n_{q_{\ast}})}{\left[  (\omega-\omega_{\mathrm{K}})^{2}%
+(\tilde{\kappa}_{0}/2+\Gamma_{1}/2)^{2}\right]  ^{2}},\label{rhor}%
\end{equation}
where $\omega_{\mathrm{K}}$ is the Kittel mode frequency of the nanowires,
$n_{L}=1/\left\{  \exp\left[  \hbar\omega_{\mathrm{K}}/(k_{B}T_{2})\right]
-1\right\}  $ and $n_{q_{\ast}}=1/\left\{  \exp\left[  \hbar\omega
_{\mathrm{K}}/(k_{B}T_{1})\right]  -1\right\}  $ are magnon numbers in left
and right wires, respectively, and $\Gamma_{1}=|g_{q_{\ast}}|^{2}/v_{q_{\ast}%
}$ is the dissipative coupling mediated by the magnons in the film. The
references signal is given by the parallel magnetization configuration of
wires and film since $g_{q_{\ast}}=0$ and the right transducer is not
affected. On the other hand, the magnons generated by a temperature gradient
via the exchange interaction at the interface or in the film, are dominantly
thermal and diffuse equally into both directions \cite{Ludo}.

Finally, we present numerical estimates for the observables. The dipolar
pumping causes additional damping $\delta\chi=\delta\tilde{\kappa}%
_{0}/(2\tilde{\omega}_{0})$ and broadening of the ferromagnetic resonance
spectrum of the nanowire. In a detector wire at a distance, the thermally
pumped magnon density $\delta\rho_{>}$ in the film injects Kittel mode magnons
$\delta\rho_{R}$. We consider a Co nanowire with width $w=70$~nm and thickness
$d=20$~nm. The magnetization $\mu_{0}\tilde{M}_{s}=1.1$~T
\cite{Yu2,Haiming_PRL}, the exchange stiffness $\tilde{\lambda}_{\mathrm{ex}%
}=3.1\times10^{-13}$~cm$^{2}$ \cite{Co_exchange} and the
Gilbert damping coefficient $\alpha_{\mathrm{Co}}=2.4\times10^{-3}$
\cite{Co_damping}. For the YIG film $s=20$~nm with magnetization $\mu_{0}%
M_{s}=0.177$~T and exchange stiffness $\lambda_{\mathrm{ex}}=3.0\times
10^{-12}$~cm$^2$ \cite{Yu2,Haiming_PRL,surface_roughness}%
. A magnetic field $\mu_{0}H_{\mathrm{app}}=0.05$~T is sufficient to switch
the film magnetizations antiparallel to that of the wire
\cite{Haiming_NC,Haiming_PRL}. The calculated additional damping of nanowire
Kittel dynamics is then $\delta\chi_{\mathrm{Co}}=3.1\times10^{-2}$, which is
one order of magnitude larger than the intrinsic one! The chiral spin Seebeck
effect is most easily resolved at low temperature. With $T_{2}=30$~K and
$T_{1}=10$~K, $\delta\rho_{>}=4\times10^{13}~\mathrm{cm}^{-2}$, $\delta
\rho_{<}=2\times10^{13}~\mathrm{cm}^{-2}$, on top of the thermal equilibrium
$\sum_{\mathbf{k}}n_{\mathbf{k}}=3\times10^{12}~\mathrm{cm}^{-2}$. The
thermally injected Kittel magnons in the detector $\delta\rho_{R}\approx10$ on
the background one $n_{q_{\ast}}\approx38$. The numbers can be strongly
increased by choosing narrower nanowires with a better chirality and placing
more than one nanowire within the spin wave propagation length, since the
signals should approximately add up. The population of tens of magnons
\cite{Yu1,Yu2} should be well within the signal to noise ratio of Brillouin
light scattering \cite{BLS_Co,BLS_review}.

\textit{Discussion}.---In conclusion, we developed a general theory of
directional (chiral) pumping of spin waves in ultrathin magnetic films. The
dipolar coupling is a relatively long-range interaction between two magnetic
bodies, which is ubiquitous in nature. At inter-magnetic interfaces it
competes with the strong, but very short-range exchange interaction, which can
easily be suppressed by inserting a non-magnetic spacer layer
\cite{CoFeB_YIG,Co_YIG,Haiming_PRL,Yu2}. The chirality generated by dipolar
interactions between magnets brings new functionalities to magnonics and
magnon spintronics \cite{logic}. Our study is closely related to the field of
chiral optics \cite{chiral_review} that focusses on electric dipoles. The
chirality of the magnetic dipolar field can be considered as the low-frequency
limit of chiral optics and plasmonics, in which retardation can be disregarded
\cite{chiral_review,science,Petersen}. We envision cross-fertilization between
optical meta-materials and magnonics, stimulating activities such as
nano-routing of magnons \cite{science,Petersen}.

\vskip0.25cm \begin{acknowledgments}
This work is financially supported by the Nederlandse Organisatie voor Wetenschappelijk Onderzoek (NWO) as well as JSPS KAKENHI Grant Nos. 26103006. We thank Prof. Haiming Yu for useful discussions.
\end{acknowledgments}

\begin{widetext}
\section{Magneto-dipolar fields}

\subsection{In-plane magnetized films}

Here we derive the dipolar field generated by spin waves in a magnetic film
with arbitrary propagation direction and ellipticity of the polarization. The
equilibrium magnetization of the film is along the $\hat{\mathbf{z}}%
$-direction. The transverse magnetization fluctuations are in general
elliptical, i.e., a superposition of the right ($m_{R}$) and left ($m_{L}$)
circular polarized components,
\[
\left(
\begin{array}
[c]{c}%
M_{x}(\mathbf{r})\\
M_{y}(\mathbf{r})
\end{array}
\right)  =\left(
\begin{array}
[c]{c}%
m_{x}^{\mathbf{k}}(x)\cos\left(  \mathbf{k}\cdot\pmb{\rho}-\omega t\right) \\
-m_{y}^{\mathbf{k}}(x)\sin\left(  \mathbf{k}\cdot\pmb{\rho}-\omega t\right)
\end{array}
\right)  =m_{R}^{\mathbf{k}}(x)\left(
\begin{array}
[c]{c}%
\cos\left(  \mathbf{k}\cdot\pmb{\rho}-\omega t\right) \\
-\sin\left(  \mathbf{k}\cdot\pmb{\rho}-\omega t\right)
\end{array}
\right)  +m_{L}^{\mathbf{k}}(x)\left(
\begin{array}
[c]{c}%
\cos\left(  \mathbf{k}\cdot\pmb{\rho}-\omega t\right) \\
\sin\left(  \mathbf{k}\cdot\pmb{\rho}-\omega t\right)
\end{array}
\right)  ,
\]
where $m_{R}^{\mathbf{k}}(x)=[m_{x}^{\mathbf{k}}(x)+m_{y}^{\mathbf{k}}(x)]/2$
and $m_{L}^{\mathbf{k}}(x)=[m_{x}^{\mathbf{k}}(x)-m_{y}^{\mathbf{k}}(x)]/2$.
This magnetization generates the dipolar field $\left(  \alpha,\beta
\in\{x,y,z\}\right)  $
\begin{equation}
h_{\beta}(\mathbf{r})=\frac{1}{4\pi}\partial_{\beta}\partial_{\alpha}\int
d\mathbf{r}^{\prime}\frac{M_{\alpha}(\mathbf{r}^{\prime})}{|\mathbf{r}%
	-\mathbf{r}^{\prime}|}. \label{hdip}%
\end{equation}
Outside the film
\begin{align}
\left(
\begin{array}
[c]{c}%
h_{x}(\mathbf{r})\\
h_{y}(\mathbf{r})\\
h_{z}(\mathbf{r})
\end{array}
\right)   &  =\frac{1}{2}\left(
\begin{array}
[c]{c}%
\left(  \left\vert \mathbf{k}\right\vert +\mathrm{sgn}(x)k_{y}\right)
\cos\left(  \mathbf{k}\cdot\pmb{\rho}-\omega t\right) \\
k_{y}\left(  \frac{k_{y}}{\left\vert \mathbf{k}\right\vert }+\mathrm{sgn}%
(x)\right)  \sin\left(  \mathbf{k}\cdot\pmb{\rho}-\omega t\right) \\
k_{z}\left(  \frac{k_{y}}{\left\vert \mathbf{k}\right\vert }+\mathrm{sgn}%
(x)\right)  \sin\left(  \mathbf{k}\cdot\pmb{\rho}-\omega t\right)
\end{array}
\right)  e^{-\left\vert \mathbf{k}\right\vert \left\vert x\right\vert }\int
dx^{\prime}m_{R}^{\mathbf{k}}\left(  x^{\prime}\right)  e^{\left\vert
	\mathbf{k}\right\vert \mathrm{sgn}(x)x^{\prime}}\nonumber\\
&  +\frac{1}{2}\left(
\begin{array}
[c]{c}%
(\left\vert \mathbf{k}\right\vert -\mathrm{sgn}(x)k_{y})\cos\left(
\mathbf{k}\cdot\pmb{\rho}-\omega t\right) \\
k_{y}\left(  -\frac{k_{y}}{\left\vert \mathbf{k}\right\vert }+\mathrm{sgn}%
(x)\right)  \sin\left(  \mathbf{k}\cdot\pmb{\rho}-\omega t\right) \\
k_{z}\left(  -\frac{k_{y}}{\left\vert \mathbf{k}\right\vert }+\mathrm{sgn}%
(x)\right)  \sin\left(  \mathbf{k}\cdot\pmb{\rho}-\omega t\right)
\end{array}
\right)  e^{-\left\vert \mathbf{k}\right\vert \left\vert x\right\vert }\int
dx^{\prime}m_{L}^{\mathbf{k}}\left(  x^{\prime}\right)  e^{\left\vert
	\mathbf{k}\right\vert \mathrm{sgn}(x)x^{\prime}},
\end{align}
where $\mathrm{sgn}(x)$ is the sign function.

The dipolar field above the film generated by a spin wave propagating normal
to the magnetization $\left(  k_{z}\rightarrow0\right)  $ reads
\cite{chiral_Yu}
\begin{align}
\left(
\begin{array}
[c]{c}%
h_{x}(\mathbf{r})\\
h_{y}(\mathbf{r})
\end{array}
\right)   &  =\frac{\left\vert k_{y}\right\vert +k_{y}}{2}\left(
\begin{array}
[c]{c}%
\cos(k_{y}y-\omega t)\\
\sin(k_{y}y-\omega t)
\end{array}
\right)  e^{-|k_{y}|x}\int dx^{\prime}m_{R}^{k_{y}}\left(  x^{\prime}\right)
e^{|k_{y}|x^{\prime}}\nonumber\\
&  +\frac{\left\vert k_{y}\right\vert -k_{y}}{2}\left(
\begin{array}
[c]{c}%
\cos(k_{y}y-\omega t)\\
-\sin(k_{y}y-\omega t)
\end{array}
\right)  e^{-|k_{y}|x}\int dx^{\prime}m_{L}^{k_{y}}\left(  x^{\prime}\right)
e^{|k_{y}|x^{\prime}},
\end{align}
while below the film
\begin{align}
\left(
\begin{array}
[c]{c}%
h_{x}(\mathbf{r})\\
h_{y}(\mathbf{r})
\end{array}
\right)   &  =\frac{\left\vert k_{y}\right\vert -k_{y}}{2}\left(
\begin{array}
[c]{c}%
\cos(k_{y}y-\omega t)\\
\sin(k_{y}y-\omega t)
\end{array}
\right)  e^{|k_{y}|x}\int dx^{\prime}m_{R}^{k_{y}}\left(  x^{\prime}\right)
e^{-\left\vert k_{y}\right\vert x^{\prime}}\nonumber\\
&  +\frac{\left\vert k_{y}\right\vert +k_{y}}{2}\left(
\begin{array}
[c]{c}%
\cos(k_{y}y-\omega t)\\
-\sin(k_{y}y-\omega t)
\end{array}
\right)  e^{|k_{y}|x}\int dx^{\prime}m_{L}^{k_{y}}\left(  x^{\prime}\right)
e^{-\left\vert k_{y}\right\vert x^{\prime}}.
\end{align}
Spin waves with right circular polarization generate a dipolar field with left
circular polarization. Right (left) propagating spin waves with $k_{y}>0$
($k_{y}<0$) only generate dipolar field above (below) the film. Spin waves
propagating parallel to the equilibrium magnetization $\left(  k_{y}%
\rightarrow0\right)  $ generate the fields
\[
\left(
\begin{array}
[c]{c}%
h_{x}(\mathbf{r})\\
h_{z}(\mathbf{r})
\end{array}
\right)  =\frac{1}{2}\left(
\begin{array}
[c]{c}%
|{k}_{z}|\cos\left(  \mathbf{k}\cdot\pmb{\rho}-\omega t\right) \\
\mathrm{sgn}\left(  x\right)  k_{z}\sin\left(  \mathbf{k}\cdot
\pmb{\rho}-\omega t\right)
\end{array}
\right)  e^{-\left\vert {k}_{z}\right\vert \left\vert x\right\vert }\int
dx^{\prime}\left(  m_{R}^{{k}_{z}}\left(  x^{\prime}\right)  +m_{L}^{{k}_{z}%
}\left(  x^{\prime}\right)  \right)  e^{|{k}_{z}|\mathrm{sgn}\left(  x\right)
	x^{\prime}}.
\]
Above the film, the dipolar field of spin waves with positive (negative)
$k_{z}$, is always left (right) circularly polarized, viz.
polarization-momentum locked. Below the film, the polarization is reversed.

\subsection{Magnetic nanowire}

\label{nanowire} Here we consider the dipolar field generated by a circularly
polarized Kittel mode and show that its Fourier components are chiral. We
consider a nanowire and its equilibrium magnetization along the $\hat
{\mathbf{z}}$ direction. The magnetic fluctuations are the real part of
\begin{equation}
\tilde{M}_{x,y}(\mathbf{r},t)=\tilde{m}_{x,y}\Theta(x)\Theta(-x+d)\Theta
(y+w/2)\Theta(-y+w/2)e^{-i\omega t},
\end{equation}
where $d$ and $w$ are the thickness and width of the nanowire. The
corresponding dipolar magnetic field
\begin{equation}
\tilde{h}_{\beta}(\mathbf{r},t)=\frac{1}{4\pi}\partial_{\beta}\partial
_{\alpha}\int\frac{\tilde{M}_{\alpha}(\mathbf{r}^{\prime},t)}{|\mathbf{r}%
	-\mathbf{r}^{\prime}|}d\mathbf{r}^{\prime}=\frac{1}{4\pi}\partial_{\beta
}\partial_{\alpha}\int dz^{\prime}\int_{0}^{d}dx^{\prime}\int_{-w/2}%
^{w/2}dy^{\prime}\frac{\tilde{m}_{\alpha}e^{-i\omega t}}{\sqrt{z^{\prime
			2}+(x-x^{\prime})^{2}+(y-y^{\prime})^{2}}}.
\end{equation}
By substituting the Coulomb integral \cite{chiral_Yu,nano_optics},
\begin{equation}
\frac{1}{\sqrt{z^{\prime2}+(x-x^{\prime})^{2}+(y-y^{\prime})^{2}}}=\frac
{1}{2\pi}\int dk_{x}dk_{y}\frac{e^{-|z^{\prime}|\sqrt{k_{x}^{2}+k_{y}^{2}}}%
}{\sqrt{k_{x}^{2}+k_{y}^{2}}}e^{ik_{x}(x-x^{\prime})+ik_{y}(y-y^{\prime})},
\end{equation}
the magnetic field below the nanowire ($x<0$) with Fourier component $k_{y}$
\begin{align}
\tilde{h}_{\beta}(k_{y},x,t)  &  =\int h_{\beta}(\mathbf{r},t)e^{-ik_{y}%
	y}dy\nonumber\\
&  =\frac{1}{2\pi}\int dk_{x}(k_{x}\tilde{m}_{x}+k_{y}\tilde{m}_{y})k_{\beta
}e^{ik_{x}x-i\omega t}\frac{1}{k_{x}^{2}+k_{y}^{2}}\frac{1}{ik_{x}%
}(1-e^{-ik_{x}d})\frac{2\sin(k_{y}w/2)}{k_{y}}.
\end{align}
Closing the contour of the $k_{x}$ integral in the lower half complex plane
\begin{equation}
\left(
\begin{array}
[c]{c}%
\tilde{h}_{x}(k_{y},x,t)\\
\tilde{h}_{y}(k_{y},x,t)
\end{array}
\right)  =-\frac{i}{4\pi}e^{\left\vert k_{y}\right\vert x}(1-e^{-\left\vert
	k_{y}\right\vert d})\frac{2\sin(k_{y}w/2)}{k_{y}\left\vert k_{y}\right\vert
}\left(
\begin{array}
[c]{cc}%
\left\vert k_{y}\right\vert  & ik_{y}\\
ik_{y} & -\left\vert k_{y}\right\vert
\end{array}
\right)  \left(
\begin{array}
[c]{c}%
\tilde{m}_{x}\\
\tilde{m}_{y}%
\end{array}
\right)  e^{-i\omega t}.
\end{equation}
A perfectly right circularly polarized wire dynamics $\left(  \tilde{m}%
_{y}=i\tilde{m}_{x}\right)  $ implies that the Fourier components of
$\tilde{\mathbf{h}}$ with $k_{y}>0$ vanish. The Fourier component with
$k_{y}<0$ is perfectly left circularly polarized $\left(  \tilde{h}%
_{y}=-i\tilde{h}_{x}\right)  $.

\section{Angle-dependent dipolar coupling}

Here we address the dependence of the coupling when the film magnetization
rotates in the film while the nanowire magnetization is kept constant. We
choose a rotated coordinate system in which the equilibrium magnetizations of
nanowire and film are $\tilde{M}_{s}\left(  0,\sin\theta,\cos\theta\right)  $
and $M_{s}\hat{\mathbf{z}}$, as shown in Fig.~\ref{coordinate}. The two
components of the dynamic magnetization in the nanowire relative to the film
magnetization are $\tilde{\mathbf{M}}_{\perp}(\mathbf{r})\parallel\left(
\tilde{\mathbf{M}}_{s}/\tilde{M}_{s}\cdot\hat{\mathbf{x}}\right)
\hat{\mathbf{x}}$ and $\tilde{\mathbf{M}}_{\parallel}(\mathbf{r}%
)\parallel\left(  \tilde{\mathbf{M}}_{s}/\tilde{M}_{s}\right)  \times
\hat{\mathbf{x}}$. 
\begin{figure}[th]
	\begin{center}
		{\includegraphics[width=9.8cm]{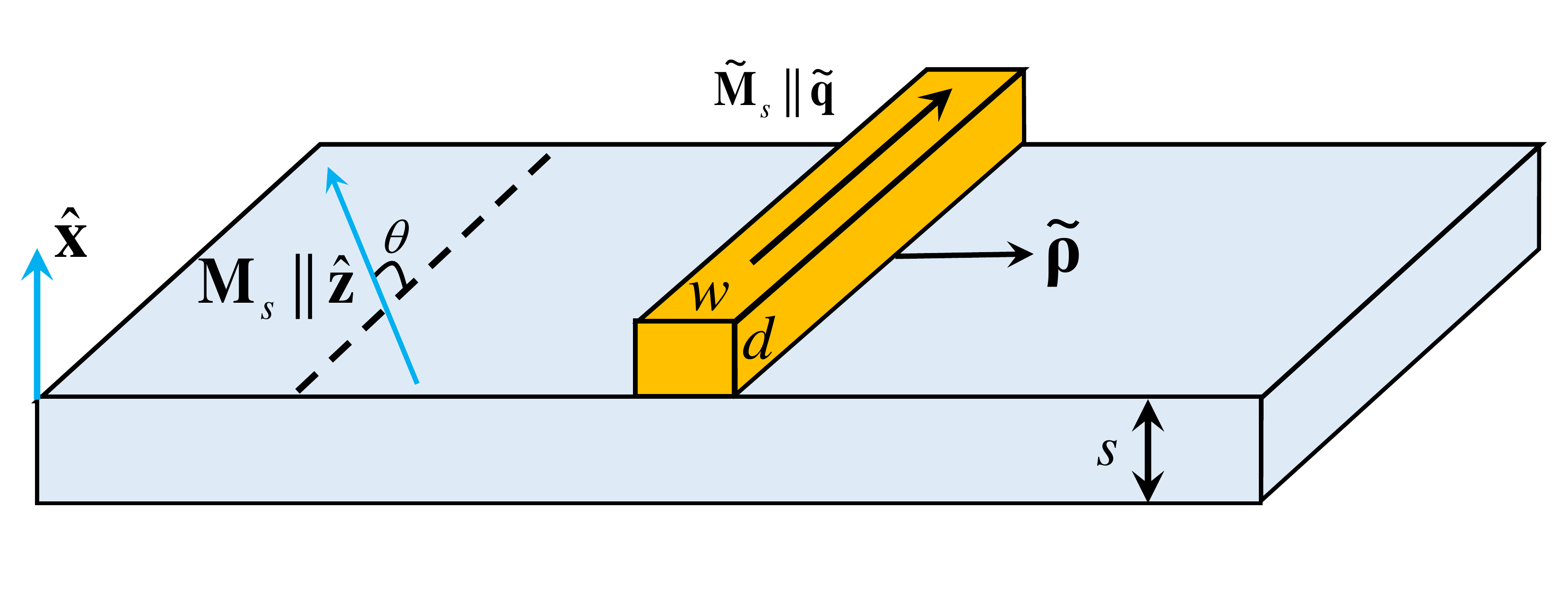}}
	\end{center}
	\caption{(Color online) Parameters and coordinate system when the
		magnetizations of film and nanowire are non-collinear.}
	\label{coordinate}%
\end{figure}The Zeeman interaction with a magnetic field $\mathbf{h}$ emitted
by the film reads
\begin{align}
H_{\mathrm{int}} &  =-\mu_{0}\int_{0}^{d}\left\{  \tilde{M}_{\perp}%
(\mathbf{r})h_{x}(\mathbf{r})+\left[  \tilde{M}_{\parallel}(\mathbf{r}%
)\cos\theta+\tilde{M}_{s}\sin\theta\right]  h_{y}(\mathbf{r})+\left[
-\tilde{M}_{\parallel}(\mathbf{r})\sin\theta+\tilde{M}_{s}\cos\theta\right]
h_{z}(\mathbf{r})\right\}  dxd\pmb{\rho}\nonumber\\
&  \rightarrow-\mu_{0}\int_{0}^{d}\left[  \tilde{M}_{\perp}(\mathbf{r}%
)h_{x}(\mathbf{r})+\tilde{M}_{\parallel}(\mathbf{r})h_{y}(\mathbf{r}%
)\cos\theta-\tilde{M}_{\parallel}(\mathbf{r})h_{z}(\mathbf{r})\sin
\theta\right]  dxd\pmb{\rho}.\label{general_Hamiltonian}%
\end{align}
The spatial integral is over the nanowire with thickness $d$ and in the second
step we disregard the fluctuating torques on the equilibrium magnetization.
The magnetization operator $\hat{M}_{\alpha}$ in the film may be expanded into
the magnon field operators $\hat{\alpha}_{\mathbf{k}}$ and $\hat{\alpha
}_{\mathbf{k}}^{\dagger}$ with Boson commutator $[\hat{\alpha}_{\mathbf{k}%
},\hat{\alpha}_{\mathbf{k}^{\prime}}^{\dagger}]=\delta_{\mathbf{k}%
	\mathbf{k}^{\prime}}$
\begin{equation}
\hat{M}_{\alpha}(\mathbf{r})=-\sqrt{2M_{s}\gamma\hbar}\sum_{\mathbf{k}}\left(
m_{\alpha}^{\mathbf{k}}(x)e^{i\mathbf{k}\cdot\pmb{\rho}}\hat{\alpha
}_{\mathbf{k}}+\overline{m_{\alpha}^{\mathbf{k}}(x)}e^{-i\mathbf{k}%
	\cdot\pmb{\rho}}\hat{\alpha}_{\mathbf{k}}^{\dagger}\right)
,\label{magnetization_film}%
\end{equation}
where $\overline{A}=A^{\ast}$ and $m_{\alpha}^{\mathbf{k}}(x)$ is the
amplitude of the spin waves over the film thickness. The magnons of nanowire
propagate with momentum $\tilde{\mathbf{q}}=\tilde{q}(\sin\theta
\hat{\mathbf{y}}+\cos\theta\hat{\mathbf{z}})=\tilde{q}\mathbf{e}_{n}$ along
the nanowire. In terms of the magnon field operators $\hat{\beta
}_{\mathbf{\tilde{q}}}$, $\hat{\beta}_{\tilde{\mathbf{q}}}^{\dagger}$ with
$[\hat{\beta}_{\mathbf{\tilde{q}}},\hat{\beta}_{\tilde{\mathbf{q}}^{\prime}%
}^{\dagger}]=\delta_{\mathbf{\tilde{q}}\tilde{\mathbf{q}}^{\prime}}$%
\begin{equation}
\hat{\tilde{M}}_{\delta}(\mathbf{r})=-\sqrt{2\tilde{M}_{s}\gamma\hbar}%
\sum_{\tilde{q}}\left(  \tilde{m}_{\delta}^{\tilde{\mathbf{q}}}(\tilde
{\pmb{\rho}})e^{i\tilde{q}\tilde{z}}\hat{\beta}_{\tilde{\mathbf{q}}}%
+\overline{\tilde{m}_{\delta}^{\tilde{\mathbf{q}}}(\tilde{\pmb{\rho}}%
	)}e^{-i\tilde{q}\tilde{z}}\hat{\beta}_{\tilde{\mathbf{q}}}^{\dagger}\right)
,\label{magnetization_wire}%
\end{equation}
where $\delta=\{\perp,\parallel\}$, $\tilde{z}=y\sin\theta+z\cos\theta$, and
$\tilde{\pmb{\rho}}=x\hat{\mathbf{x}}+\tilde{y}(\cos\theta\hat{\mathbf{y}%
}-\sin\theta\hat{\mathbf{z}})$ is a vector in the nanowire cross section with
$-w/2\leq\tilde{y}\leq w/2$ and $0\leq x\leq d$.

Using the dipolar field Eq. (\ref{hdip})
\begin{equation}
\hat{h}_{\beta}(\mathbf{r})=\frac{1}{4\pi}\partial_{\beta}\partial_{\alpha
}\int d\mathbf{r}^{\prime}\frac{\hat{M}_{\alpha}(\mathbf{r}^{\prime}%
	)}{|\mathbf{r}-\mathbf{r}^{\prime}|},
\end{equation}
and substituting Eqs.~(\ref{magnetization_film}) and (\ref{magnetization_wire}%
) into Eq.~(\ref{general_Hamiltonian}) yields
\begin{equation}
\hat{H}_{\mathrm{int}}=\sum_{\mathbf{k}}\left(  g_{\mathbf{k}}\hat{\alpha
}_{\mathbf{k}}^{\dagger}\hat{\beta}_{k_{\parallel}\mathbf{e}_{n}%
}+\mathrm{h.c.}\right)  ,
\end{equation}
where $k_{\parallel}=k_{y}\sin\theta+k_{z}\cos\theta$, the coupling constant
\[
g_{\mathbf{k}}=-2\mu_{0}\gamma\hbar\sqrt{\tilde{M}_{s}M_{s}}\frac{1}{k_{\perp
}}\sin\left(  \frac{k_{\perp}w}{2}\right)  \int_{0}^{d}dx\int_{-s}%
^{0}dx^{\prime-(x-x^{\prime})|\mathbf{k}|}\left(  \overline{m_{x}^{\mathbf{k}%
	}(x^{\prime})},\overline{m_{y}^{\mathbf{k}}(x^{\prime})}\right)  \left(
\begin{array}
[c]{cc}%
|\mathbf{k}| & ik_{\perp}\\
ik_{y} & -\frac{k_{y}k_{\perp}}{|\mathbf{k}|}%
\end{array}
\right)  \left(
\begin{array}
[c]{c}%
\tilde{m}_{\perp}^{k_{\parallel}\mathbf{e}_{n}}(x)\\
\tilde{m}_{\parallel}^{k_{\parallel}\mathbf{e}_{n}}(x)
\end{array}
\right)  ,
\]
and $k_{\perp}=-k_{z}\sin\theta+k_{y}\cos\theta$. For thin films the
magnetization is constant over the film $\left(  s\right)  $ and nanowire
$\left(  d\right)  $ thickness and
\[
g_{\mathbf{k}}\rightarrow-2\mu_{0}\gamma\hbar\sqrt{\tilde{M}_{s}M_{s}}\frac
{1}{k_{\perp}|\mathbf{k}|^{2}}\sin\left(  \frac{k_{\perp}w}{2}\right)  \left(
1-e^{-|\mathbf{k}|d}\right)  \left(  1-e^{-|\mathbf{k}|s}\right)  \left(
\overline{m_{x}^{\mathbf{k}}},\overline{m_{y}^{\mathbf{k}}}\right)  \left(
\begin{array}
[c]{cc}%
|\mathbf{k}| & ik_{\perp}\\
ik_{y} & -\frac{k_{y}k_{\perp}}{|\mathbf{k}|}%
\end{array}
\right)  \left(
\begin{array}
[c]{c}%
\tilde{m}_{\perp}^{k_{\parallel}\mathbf{e}_{n}}\\
\tilde{m}_{\parallel}^{k_{\parallel}\mathbf{e}_{n}}%
\end{array}
\right)  .
\]
The normalized magnon amplitudes of exchange spin waves in the film
\cite{Walker,surface_roughness,chiral_Yu}
\begin{equation}
m_{y}^{\mathbf{k}}=im_{x}^{\mathbf{k}}=i\sqrt{1/(4s)}, \label{exchange_waves}%
\end{equation}
and those in the nanowire are
\begin{equation}
\tilde{m}_{\perp}^{k_{\parallel}\mathbf{e}_{n}}=\sqrt{\frac{1}{4\mathcal{D}%
		(k_{\parallel})wd}},~~~~\tilde{m}_{\parallel}^{k_{\parallel}\mathbf{e}_{n}%
}=i\sqrt{\frac{\mathcal{D}(k_{\parallel})}{4wd}}, \label{nanowire_waves}%
\end{equation}
where
\begin{equation}
\mathcal{D}(k_{\parallel})=\sqrt{\frac{H_{\mathrm{app}}+N_{xx}\tilde{M}%
		_{s}+\tilde{\lambda}_{\mathrm{ex}}k_{\parallel}^{2}\tilde{M}_{s}%
	}{H_{\mathrm{app}}+N_{yy}\tilde{M}_{s}+\tilde{\lambda}_{\mathrm{ex}%
		}k_{\parallel}^{2}\tilde{M}_{s}}}.
\end{equation}
$H_{\mathrm{app}}$ and $\tilde{\lambda}_{\mathrm{ex}}$ are the applied
magnetic field and the exchange stiffness of the nanowire, respectively. The
demagnetization factors are estimated to be $N_{xx}\simeq w/(d+w)$ and
$N_{yy}=d/(d+w)$ \cite{chiral_Yu} also govern the Kittel mode frequency
\begin{equation}
\omega_{\mathrm{K}}=\mu_{0}\gamma\sqrt{(H_{\mathrm{app}}+N_{yy}\tilde{M}%
	_{s})(H_{\mathrm{app}}+N_{xx}\tilde{M}_{s})}. \label{omegaK}%
\end{equation}

We can now discuss special configurations.\newline(i) When magnetizations are
antiparallel, $\theta=\pi$, $k_{\parallel}=-k_{z}$, $k_{\perp}=-k_{y}$, and
$\mathbf{e}_{n}=-\hat{\mathbf{z}}$. The coupling strength
\begin{align}
g_{\mathbf{k}}^{\parallel}  &  \rightarrow-2\mu_{0}\gamma\hbar\sqrt{\tilde
	{M}_{s}M_{s}}\frac{1}{k_{y}|\mathbf{k}|^{2}}\sin\left(  \frac{k_{y}w}%
{2}\right)  \left(  1-e^{-|\mathbf{k}|d}\right)  \left(  1-e^{-|\mathbf{k}%
	|s}\right)  \left(  \overline{m_{x}^{\mathbf{k}}},\overline{m_{y}^{\mathbf{k}%
}}\right)  \left(
\begin{array}
[c]{cc}%
|\mathbf{k}| & -ik_{y}\\
ik_{y} & \frac{k_{y}^{2}}{|\mathbf{k}|}%
\end{array}
\right)  \left(
\begin{array}
[c]{c}%
\tilde{m}_{\perp}^{k_{z}\hat{\mathbf{z}}}\\
\tilde{m}_{\parallel}^{k_{z}\hat{\mathbf{z}}}%
\end{array}
\right)  .
\end{align}
In the notation of the main text, $\tilde{m}_{\perp}^{k_{z}\hat{\mathbf{z}}%
}=\tilde{m}_{x}^{(k_{z})}$ and $\tilde{m}_{\parallel}^{k_{z}\hat{\mathbf{z}}%
}=-\tilde{m}_{y}^{(k_{z})}$. When both spin waves in the film and nanowire are
circularly polarized the chirality is perfect and the coupling strength is
maximized.\newline(ii) When magnetizations are normal to each other,
$\theta=\pi/2$, $k_{\parallel}=k_{y}$, $k_{\perp}=-k_{z}$, $\mathbf{e}%
_{n}=\hat{\mathbf{y}}$, and
\begin{align}
g_{\mathbf{k}}^{\perp}  &  \rightarrow-2\mu_{0}\gamma\hbar\sqrt{\tilde{M}%
	_{s}M_{s}}\frac{1}{k_{z}|\mathbf{k}|^{2}}\sin\left(  \frac{k_{z}w}{2}\right)
\left(  1-e^{-|\mathbf{k}|d}\right)  \left(  1-e^{-|\mathbf{k}|s}\right)
\left(  \overline{m_{x}^{\mathbf{k}}},\overline{m_{y}^{\mathbf{k}}}\right)
\left(
\begin{array}
[c]{cc}%
|\mathbf{k}| & -ik_{z}\\
ik_{y} & \frac{k_{y}k_{z}}{|\mathbf{k}|}%
\end{array}
\right)  \left(
\begin{array}
[c]{c}%
\tilde{m}_{\perp}^{k_{y}\hat{\mathbf{y}}}\\
\tilde{m}_{\parallel}^{k_{y}\hat{\mathbf{y}}}%
\end{array}
\right)  .
\end{align}
The coupling to travelling waves in the nanowire with finite $k_{y}$ is not
perfectly chiral, even for the circularly polarized spin waves in the film and
nanowire, but still directional, depending on $k_{y}/k_{z}$.\newline(iii) In
the limit of coherent excitation of only the Kittel mode $k_{\parallel}=0$,
but at arbitrary angle $k_{y}=k_{\perp}\cos\theta$,
\begin{align}
g_{k_{\perp}}^{\mathrm{K}}  &  \rightarrow-2\mu_{0}\gamma\hbar\sqrt{\tilde
	{M}_{s}M_{s}}\frac{1}{k_{\perp}^{3}}\sin\left(  \frac{k_{\perp}w}{2}\right)
\left(  1-e^{-|k_{\perp}|d}\right)  \left(  1-e^{-|k_{\perp}|s}\right)
\left(  \overline{m_{x}^{\mathbf{k}}},\overline{m_{y}^{\mathbf{k}}}\cos
\theta\right)  \left(
\begin{array}
[c]{cc}%
|{k}_{\perp}| & ik_{\perp}\\
ik_{\perp} & -|k_{\perp}|
\end{array}
\right)  \left(
\begin{array}
[c]{c}%
\tilde{m}_{\perp}^{(0)}\\
\tilde{m}_{\parallel}^{(0)}%
\end{array}
\right)  .
\end{align}
The Kittel mode in the nanowire with right circular polarization couples with
the spin waves propagating perpendicular to the nanowire with perfect
chirality (Sec.~\ref{nanowire}). In general, the Kittel mode in the nanowire
is elliptic; the chirality can then be tuned by the angle $\theta$. In
particular, the ellipticity leads to a \textquotedblleft
magic\textquotedblright\ angle $\theta_{c}$ at which the chirality of the
(nonzero) coupling vanishes. Using $\left\vert g_{k_{\perp}}^{\mathrm{K}%
}\right\vert =|g_{-k_{\perp}}^{\mathrm{K}}|$ and assuming pure exchange spin
waves in the film [Eq.~(\ref{exchange_waves})],
\[
\cos\theta_{c}=-i\tilde{m}_{\parallel}^{(0)}/\tilde{m}_{\perp}^{(0)}%
~~\mathrm{or}~~i\tilde{m}_{\perp}^{(0)}/\tilde{m}_{\parallel}^{(0)}.
\]
In the limit of small applied magnetic fields, Eq.~(\ref{nanowire_waves})
yields $\tilde{m}_{\perp}^{(0)}\simeq\sqrt{1/(4w\sqrt{wd})}$ and $\tilde
{m}_{\parallel}^{(0)}\simeq i\sqrt{1/(4d\sqrt{wd})}$. With $w>d$, the critical
angle is governed by the aspect ratio with $\cos\theta_{c}\simeq\sqrt{d/w}$.
When $d\rightarrow w\ $the Kittel mode is circularly polarized and the
chirality vanishes with the coupling constant when approaching the parallel configuration.

\section{Linear response theory of chiral magnon excitation}

The coherent excitation of a magnetization by a proximity magnetic transducer
can alternatively be formulated by linear response theory
\cite{Mahan,pumping_linear}. The excited magnetization in the film can be
expressed by time-dependent perturbation theory as:
\begin{equation}
{M}_{\alpha}(x,\pmb{\rho},t)=-i\int_{-\infty}^{t}dt^{\prime}\left\langle
\left[  \hat{M}_{\alpha}(x,\pmb{\rho},t),\hat{H}_{\mathrm{int}}(t^{\prime
})\right]  \right\rangle .
\end{equation}
In terms of the retarded spin susceptibility
\begin{equation}
\chi_{\alpha\delta}(x,x^{\prime};\pmb{\rho}-\pmb{\rho}^{\prime};t-t^{\prime
})=i\Theta(t-t^{\prime})\left\langle \left[  \hat{S}_{\alpha}%
(x,\pmb{\rho},t),\hat{S}_{\delta}(x^{\prime},\pmb{\rho}^{\prime},t^{\prime
})\right]  \right\rangle ,
\end{equation}
where $\hat{S}_{\alpha}=-\hat{M}_{\alpha}/(\gamma\hbar)$ is the spin
operator,
\[
{M}_{\alpha}(x,\pmb{\rho},t)=\mu_{0}(\gamma\hbar)^{2}\sum_{\mathbf{k}}%
\int_{-\infty}^{\infty}dt^{\prime}\int_{0}^{d}d\tilde{x}d\tilde{\pmb{\rho}}%
\int_{-s}^{0}dx^{\prime}\tilde{M}_{\beta}(\tilde{x},\tilde{\pmb{\rho}}%
,t^{\prime})G_{\beta\xi}(-\mathbf{k},\tilde{x}-x^{\prime})\chi_{\alpha\xi
}\left(  x,x^{\prime};\mathbf{k};t-t^{\prime}\right)  e^{i\mathbf{k}%
	\cdot(\pmb{\rho}-\tilde{\pmb{\rho}})}.
\]
Here $\tilde{\mathbf{M}}$ is the magnetization of the magnetic transducer.
With $\tilde{x}>x^{\prime}$, the Green-function tensor reads
\begin{equation}
G(-\mathbf{k},\tilde{x}-x^{\prime})=\frac{e^{-|\tilde{x}-x^{\prime
		}||\mathbf{k}|}}{2}\left(
\begin{array}
[c]{ccc}%
|\mathbf{k}| & ik_{y} & ik_{z}\\
ik_{y} & -k_{y}^{2}/|\mathbf{k}| & -k_{y}k_{z}/|\mathbf{k}|\\
ik_{z} & -k_{y}k_{z}/|\mathbf{k}| & -k_{z}^{2}/|\mathbf{k}|
\end{array}
\right)  .
\end{equation}
In terms of eigenmodes $m_{\alpha}^{\mathbf{k}}(x)e^{i\mathbf{k}%
	\cdot\pmb{\rho}}$ and their frequency $\omega_{\mathbf{k}}$,
\begin{equation}
\chi_{\alpha\xi}(x,x^{\prime};\mathbf{k};\omega)=-\frac{2M_{s}}{\gamma\hbar
}m_{\alpha}^{\mathbf{k}}(x)\overline{m_{\xi}^{\mathbf{k}}(x)}\frac{1}%
{\omega-\omega_{\mathbf{k}}+i0_{+}}%
\end{equation}
is the spin susceptibility in momentum-frequency space. The excited
magnetization is
\begin{align}
M_{\alpha}(x,\pmb{\rho},t)  &  =-2\mu_{0}M_{s}\gamma\hbar\sum_{\mathbf{k}}%
\int_{-\infty}^{\infty}dt^{\prime}\int_{0}^{d}d\tilde{x}d\tilde{\pmb{\rho}}%
\int_{-s}^{0}dx^{\prime}\int\frac{d\omega}{2\pi}e^{-i\omega(t-t^{\prime
	})+i\mathbf{k}\cdot(\pmb{\rho}-\tilde{\pmb{\rho}})}\frac{1}{\omega
	-\omega_{\mathbf{k}}+i0_{+}}\nonumber\\
&  \times m_{\alpha}^{\mathbf{k}}(x)\tilde{M}_{\beta}(\tilde{x},\tilde
{\pmb{\rho}},t^{\prime})G_{\beta\xi}(-\mathbf{k},\tilde{x}-x^{\prime
})\overline{m_{\xi}^{\mathbf{k}}(x)}.
\end{align}

Under steady-state resonant microwave excitation of the Kittel mode Eq.
(A\ref{omegaK})
\begin{equation}
\tilde{M}_{\beta}(\tilde{x},\tilde{\pmb{\rho}},t^{\prime})\approx\tilde
{M}_{\beta}(\tilde{x},\tilde{\pmb{\rho}},t)e^{i\omega_{\mathrm{K}}%
	(t-t^{\prime})},
\end{equation}
the film magnetization becomes
\[
M_{\alpha}(x,\pmb{\rho},t)=-2\mu_{0}M_{s}\gamma\hbar\sum_{\mathbf{k}}\int%
_{0}^{d}d\tilde{x}d\tilde{\pmb{\rho}}\int_{-s}^{0}dx^{\prime}e^{i\mathbf{k}%
	\cdot(\pmb{\rho}-\tilde{\pmb{\rho}})}\frac{1}{\omega_{\mathrm{K}}%
	-\omega_{\mathbf{k}}+i0_{+}}m_{\alpha}^{\mathbf{k}}(x)\tilde{M}_{\beta}%
(\tilde{x},\tilde{\pmb{\rho}},t)G_{\beta\xi}(-\mathbf{k},\tilde{x}-x^{\prime
})\overline{m_{\xi}^{\mathbf{k}}(x)}.
\]
When nanowire and equilibrium magnetizations are parallel to $\mathbf{\hat{z}%
}$, the momentum integral in
\begin{align}
M_{\alpha}(x,y,t)  &  =-2\mu_{0}M_{s}\gamma\hbar\int\frac{dk_{y}}{2\pi}%
\int_{0}^{d}d\tilde{x}d\tilde{y}\int_{-s}^{0}dx^{\prime}e^{ik_{y}(y-\tilde
	{y})}\frac{1}{\omega_{\mathrm{K}}-\omega_{k_{y}}+i0_{+}}\nonumber\\
&  \times m_{\alpha}^{k_{y}}(x)M_{\beta}(\tilde{x},\tilde{y},t)G_{\beta\xi
}(-k_{y},\tilde{x}-x^{\prime})\overline{m_{\xi}^{k_{y}}(x)}%
\end{align}
can be evaluated by contours in the complex plane. The zeros of the
denominator $\omega_{\mathrm{K}}-\omega_{k_{y}}+i0_{+}$ generate two
singularities at $k_{\pm}=\pm(k_{\ast}+i0_{+})$ with $k_{\ast}>0$, so $k_{+}$
and $k_{-}$ lie in the upper and lower half planes, respectively. When
$y>\tilde{y}$ the contour should be closed in the upper half plane and
\begin{equation}
M_{\alpha}^{>}(x,y,t)=2i\mu_{0}M_{s}\gamma\hbar\frac{1}{v_{k_{\ast}}}\int%
_{0}^{d}d\tilde{x}d\tilde{y}\int_{-s}^{0}dx^{\prime}e^{ik_{\ast}(y-\tilde{y}%
	)}m_{\alpha}^{k_{\ast}}(x)M_{\beta}(\tilde{x},\tilde{y},t)G_{\beta\xi
}(-k_{\ast},\tilde{x}-x^{\prime})\overline{m_{\xi}^{k_{\ast}}(x)},
\end{equation}
where $v_{k_{\ast}}=\left.  \partial\omega_{k}/\partial k\right\vert
_{k=k_{\ast}}$ is the spin wave group velocity. A small or zero group velocity
implies a large density of states and excitation efficiency. When $y<\tilde
{y}$,
\begin{equation}
M_{\alpha}^{<}(x,y,t)=2i\mu_{0}M_{s}\gamma\hbar\frac{1}{v_{k_{\ast}}}\int%
_{0}^{d}d\tilde{x}d\tilde{y}\int_{-s}^{0}dx^{\prime}e^{-ik_{\ast}(y-\tilde
	{y})}m_{\alpha}^{-k_{\ast}}(x)M_{\beta}(\tilde{x},\tilde{y},t)G_{\beta\xi
}(k_{\ast},\tilde{x}-x^{\prime})\overline{m_{\xi}^{-k_{\ast}}(x)}.
\end{equation}
When the spin waves in the film are circularly polarized with $m_{y}=im_{x}$,
\begin{equation}
G_{\beta\xi}(-k_{\ast},\tilde{x}-x^{\prime})\overline{m_{\xi}^{k_{\ast}}%
	(x)}\longrightarrow\frac{e^{-|\tilde{x}-x^{\prime}||{k}_{\ast}|}}{2}\left(
\begin{array}
[c]{cc}%
k_{\ast} & ik_{\ast}\\
ik_{\ast} & -k_{\ast}%
\end{array}
\right)  \left(
\begin{array}
[c]{c}%
m_{x}\\
-im_{x}%
\end{array}
\right)  =\left(
\begin{array}
[c]{c}%
0\\
0
\end{array}
\right)  ,
\end{equation}
leading to zero $M_{\alpha}^{<}(x,y,t)$, but finite $M_{\alpha}^{>}(x,y,t)$.
So the nanowire can only excite spin waves with positive momentum. Also,
energy and momentum is injected into only half of the film with $y>\tilde{y}$
. This \textquotedblleft spatial chirality\textquotedblright\ persists in the
limit of vanishing dissipation and is a consequence of the causality or retardation.

\section{Scattering matrix of microwave photons}

The magnetic order in two nanowires located at $\mathbf{r}_{1}=R_{1}%
\hat{\mathbf{y}}$ and $\mathbf{r}_{2}=R_{2}\hat{\mathbf{y}}$ may act as
transducers for microwaves that are emitted or detected by local microwave
antennas as well as excite and detect magnons in the film. We are interested
in the observable---the scattering matrix of the microwaves with excitation
(input) at $R_{1}$ and the detection (output) at $R_{2}$, which can be
formulated by the input-output theory \cite{S_input_output1,S_input_output2}.
When the local magnon states at $R_{1}$ and $R_{2}$ are expressed by the
operators $\hat{m}_{L}$ and $\hat{m}_{R}$, respectively, this leads to the
equations of motion of the coupled nanowires and the film
\begin{align}
\frac{d\hat{m}_{L}}{dt} &  =-i\omega_{\mathrm{K}}\hat{m}_{L}(t)-i\sum_{q}%
g_{q}e^{iqR_{1}}\hat{\alpha}_{q}(t)-\left(  \frac{\kappa_{L}}{2}+\frac
{\kappa_{p,L}}{2}\right)  \hat{m}_{L}(t)-\sqrt{\kappa_{p,L}}\hat
{p}_{\mathrm{in}}^{(L)}(t),\nonumber\\
\frac{d\hat{m}_{R}}{dt} &  =-i\omega_{\mathrm{K}}\hat{m}_{R}(t)-i\sum_{q}%
g_{q}e^{iqR_{2}}\hat{\alpha}_{q}(t)-\frac{\kappa_{R}}{2}\hat{m}_{R}%
(t),\nonumber\\
\frac{d\hat{\alpha}_{q}}{dt} &  =-i\omega_{q}\hat{\alpha}_{q}(t)-ig_{q}%
e^{-iqR_{1}}\hat{m}_{L}(t)-ig_{q}e^{-iqR_{2}}\hat{m}_{R}(t)-\frac{\kappa_{q}%
}{2}\hat{\alpha}_{q}(t).
\end{align}
Here, $\kappa_{L}$ and $\kappa_{R}$ are the intrinsic damping of the Kittel
modes in the left and right nanowires, respectively, $\kappa_{p,L}$ is the
additional radiative damping induced by the microwave photons $\hat
{p}_{\mathrm{in}}^{(L)}$, i.e. the coupling of the left nanowire with the
microwave source, and $\kappa_{q}$ denotes the intrinsic (Gilbert) damping of
magnons in the films. In frequency space:
\begin{align}
\hat{\alpha}_{q}(\omega) &  =g_{q}G_{q}\left(  \omega\right)  \left[
e^{-iqR_{1}}\hat{m}_{L}(\omega)+e^{-iqR_{2}}\hat{m}_{R}(\omega)\right]
,\nonumber\\
\hat{m}_{R}(\omega) &  =\frac{-i\sum_{q}g_{q}^{2}G_{q}\left(  \omega\right)
	e^{iq(R_{2}-R_{1})}}{-i(\omega-\omega_{\mathrm{K}})+\kappa_{R}/2+i\sum
	_{q}g_{q}^{2}G_{q}\left(  \omega\right)  }\hat{m}_{L}(\omega),\nonumber\\
\hat{m}_{L}(\omega) &  =\frac{-\sqrt{\kappa_{p,L}}}{-i(\omega-\omega
	_{\mathrm{K}})+(\kappa_{L}+\kappa_{p,L})/2+i\sum_{q}g_{q}^{2}G_{q}\left(
	\omega\right)  -f(\omega)}\hat{p}_{\mathrm{in}}^{(L)}(\omega),
\end{align}
where $G_{q}\left(  \omega\right)  =\left[  (\omega-\omega_{q})+i\kappa
_{q}/2\right]  ^{-1}$ and
\begin{equation}
f(\omega)\equiv-\frac{\left(  \sum_{q}g_{q}^{2}G_{q}\left(  \omega\right)
	e^{iq(R_{1}-R_{2})}\right)  \left(  \sum_{q}g_{q}^{2}G_{q}\left(
	\omega\right)  e^{iq(R_{2}-R_{1})}\right)  }{-i(\omega-\omega_{\mathrm{K}%
	})+\kappa_{R}/2+i\sum_{q}g_{q}^{2}G_{q}\left(  \omega\right)  }.
\end{equation}
For perfect chiral coupling $f(\omega)$ vanishes by the absence of
back-action. The excitation of the left nanowire propagates to the right
nanowire by the spin waves in the film. The microwave output of the left and
right nanowires inductively detected by coplanar wave guides are denoted by
$\hat{p}_{\mathrm{out}}^{(L)}(\omega)$ and $\hat{p}_{\mathrm{out}}%
^{(R)}(\omega)$ with input-output relations
\cite{S_input_output1,S_input_output2}
\begin{align}
\hat{p}_{\mathrm{out}}^{(L)}(\omega) &  =p_{\mathrm{in}}^{(L)}(\omega
)+\sqrt{\kappa_{p,L}}\hat{m}_{L}(\omega),\nonumber\\
\hat{p}_{\mathrm{out}}^{(R)}(\omega) &  =\sqrt{\kappa_{p,R}}\hat{m}_{R}%
(\omega),
\end{align}
where $\kappa_{p,R}$ is the additional radiative damping induced by the
detector. Therefore, the elements in the microwave scattering matrix, i.e.,
microwave reflection $\left(  S_{11}\right)  $ and transmission $\left(
S_{21}\right)  $ amplitudes become
\begin{align}
S_{11}(\omega) &  \equiv\frac{\hat{p}_{\mathrm{out}}^{(L)}}{\hat
	{p}_{\mathrm{in}}^{(L)}}=1-\frac{\kappa_{p,L}}{-i(\omega-\omega_{\mathrm{K}%
	})+(\kappa_{L}+\kappa_{p,L})/2+i\sum_{q}g_{q}^{2}G_{q}\left(  \omega\right)
	-f(\omega)},\nonumber\\
S_{21}(\omega) &  \equiv\frac{\hat{p}_{\mathrm{out}}^{(R)}}{\hat
	{p}_{\mathrm{in}}^{(L)}}=\left[  1-S_{11}(\omega)\right]  \sqrt{\frac
	{\kappa_{p,R}}{\kappa_{p,L}}}\frac{i\sum_{q}g_{q}^{2}G_{q}\left(
	\omega\right)  e^{iq(R_{2}-R_{1})}}{-i(\omega-\omega_{\mathrm{K}})+\kappa
	_{R}/2+i\sum_{q}g_{q}^{2}G_{q}\left(  \omega\right)  }.\label{S}%
\end{align}
The real parts of $S_{11}$ and $S_{12}$ are illustrated in
Fig.~\ref{transmission} when the magnetizations of nanowire and film are
antiparallel. \begin{figure}[th]
	\begin{center}
		{\includegraphics[width=8.0cm]{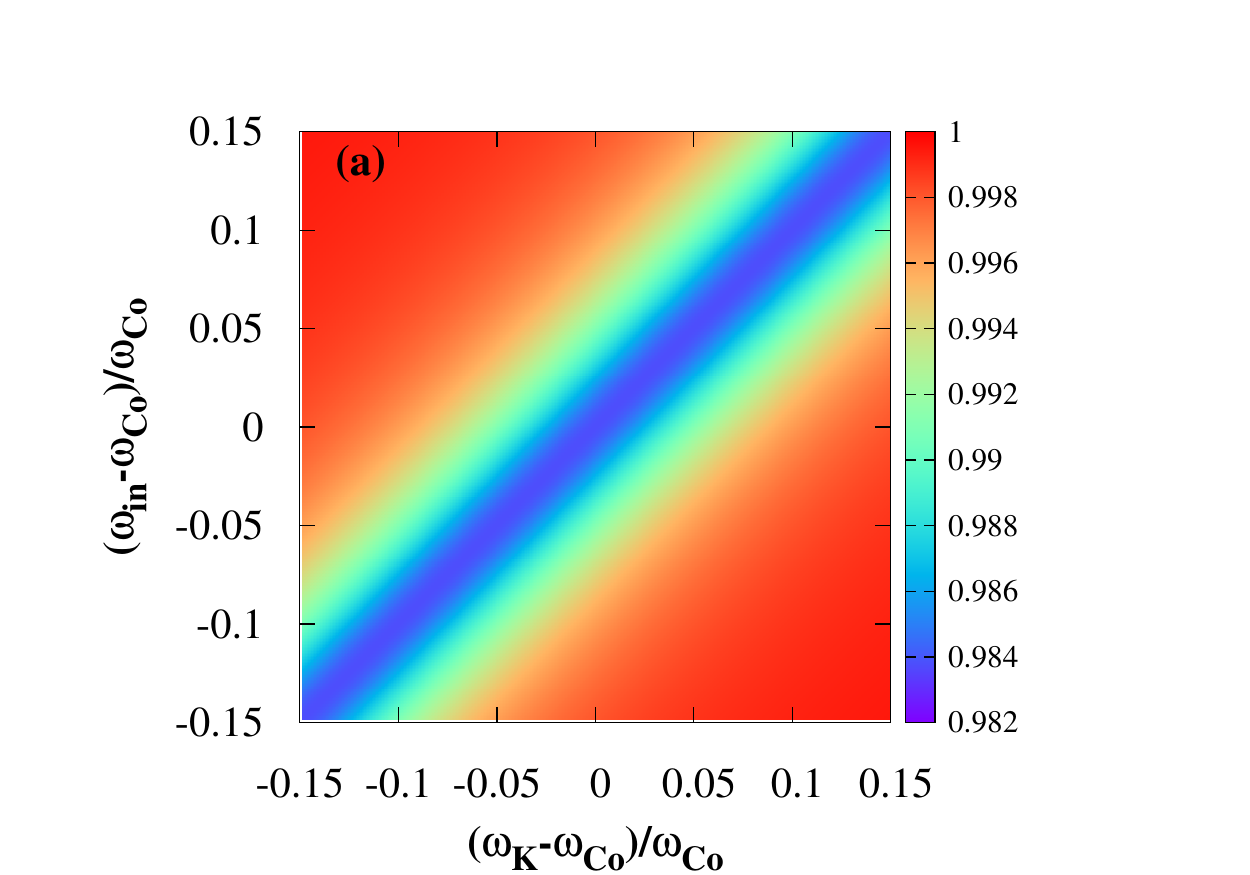}} \hspace{-0.4cm}
		{\includegraphics[width=8.0cm]{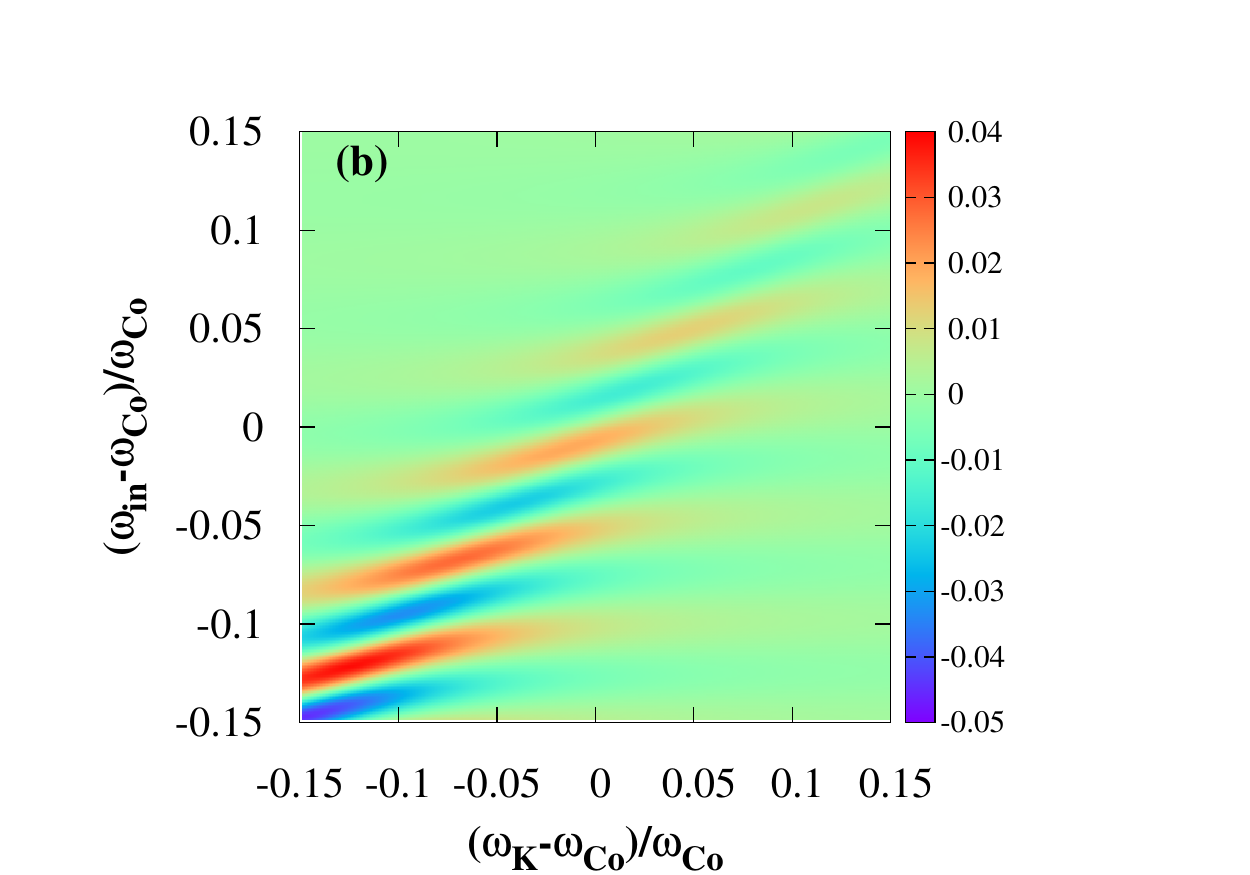}}
	\end{center}
	\caption{(Color online) Reflection $\operatorname{Re}(S_{11})$ [(a)] and
		transmission $\operatorname{Re}(S_{12})$ [(b)] amplitudes of microwaves, Eq.
		(A\ref{S}) between two magnetic nanowires on a magnetic film. $\omega
		_{\mathrm{Co}}$ is the Kittel mode frequency Eq.~(\ref{omegaK}) of the cobalt
		nanowire at a small applied field ($H_{\mathrm{app}}=0.05$~T) that fixes an
		antiparallel magnetizations, $\omega_{\mathrm{in}}$ is the frequency of the
		input microwaves, and $\omega_{\mathrm{K}}$ is the Kittel mode frequency as a
		function of an applied field $H_{\mathrm{app}}.$ The radiative coupling of
		both nanowires $\kappa_{p}/(2\pi)=10$~MHz while other parameter values are
		listed in the main text. }%
	\label{transmission}%
\end{figure} the interference patterns on the Kittel resonance of cobalt
nanowire in Fig.~\ref{transmission}(b) reflect the interaction between
nanowires and film. The phase factor $e^{ik(R_{1}-R_{2})}$ in Eq.~(\ref{S})
provides peaks and dips when the resonant momentum $k$ is modulated. These
patterns are not caused by spin wave interference in the film since in our
model the nanowires cannot reflect spin waves.

\section{Dipolar non-local spin Seebeck effect}

We consider two identical transducers, with a magnetic nanowire at
$\mathbf{r}_{2}=R_{2}\hat{\mathbf{y}}$ that detects thermally injected magnons
by a nanowire at $\mathbf{r}_{1}=R_{1}\hat{\mathbf{y}}$ with $R_{1}<R_{2}$
mediated by the dipolar interaction only. For simplicity, we consider only the
Kittel modes in the wires, which is a good approximation at low temperatures
at which higher modes are frozen out. The contribution by higher modes with
large wave numbers $k$ is disregarded because the dipolar coupling is
exponentially suppressed $\sim e^{-kx}$. The coupling strength $|g_{\mathbf{k}%
}|$ in Fig.~\ref{coupling} illustrates that magnons with wavelength around
half of the nanowire width ($\pi/w=0.045$~nm$^{-1}$) dominate the coupling.
Thermal pumping from other than the Kittel mode can be disregarded even at
elevated temperatures. Furthermore, the spin current in the film is dominated
by spin waves with small momentum and long mean-free paths, so in the
following we may disregard the effects of magnon-magnon and magnon-phonon
interactions that otherwise render magnon transport phenomena diffuse
\cite{Ludo}. The narrow-band thermal injection also favors the inductive
detection of the injected spin current pursued here, rather than by the
inverse spin Hall effect with heavy metal contacts.

The equation of motions of the Kittel modes in the nanowire and film spin
waves in the coupled system read
\begin{align}
\frac{d\hat{m}_{L}}{dt}  &  =-i\omega_{\mathrm{K}}\hat{m}_{L}-\sum_{q}%
ig_{q}^{\ast}e^{iqR_{1}}\hat{\alpha}_{q}-\frac{\kappa}{2}\hat{m}_{L}%
-\sqrt{\kappa}\hat{N}_{L},\nonumber\\
\frac{d\hat{m}_{R}}{dt}  &  =-i\omega_{\mathrm{K}}\hat{m}_{R}-\sum_{q}%
ig_{q}^{\ast}e^{iqR_{2}}\hat{\alpha}_{q}-\frac{\kappa}{2}\hat{m}_{R}%
-\sqrt{\kappa}\hat{N}_{R},\nonumber\\
\frac{d\hat{\alpha}_{q}}{dt}  &  =-i\omega_{q}\hat{\alpha}_{q}-ig_{q}%
e^{-iqR_{1}}\hat{m}_{L}-ig_{q}e^{-iqR_{2}}\hat{m}_{R}-\frac{\kappa_{q}}{2}%
\hat{\alpha}_{q}-\sqrt{\kappa_{q}}\hat{N}_{q},
\end{align}
where $\kappa$ is caused by the same Gilbert damping in both nanowires, and
$\hat{N}_{L}$ and $\hat{N}_{R}$ represent the thermal noise in the left and
right nanowires, with $\langle{\hat{N}_{\eta}^{\dagger}(t)\hat{N}%
	_{\eta^{\prime}}(t^{\prime})}\rangle=n_{\eta}\delta(t-t^{\prime})\delta
_{\eta\eta^{\prime}}$. Here, $\eta\in\{L,R\}$ and $n_{\eta}=1/\left\{
\exp\left[  \hbar\omega_{\mathrm{K}}/(k_{B}T_{\eta})\right]  -1\right\}  $ and
$T_{R}$ is also the film temperature. Integrating out the spin-wave modes in
the film, we obtain equations for dissipatively coupled nanowires. In
frequency space,
\begin{align}
\left(  -i(\omega-\omega_{\mathrm{K}})+\frac{\kappa}{2}+\frac{\Gamma
	_{1}+\Gamma_{2}}{2}\right)  \hat{m}_{L}(\omega)+\Gamma_{2}e^{iq_{\ast}%
	|R_{2}-R_{1}|}\hat{m}_{R}(\omega)  &  =\sum_{q}ig_{q}^{\ast}e^{iqR_{1}}%
\sqrt{\kappa_{q}}G_{q}(\omega)\hat{N}_{q}(\omega)-\sqrt{\kappa}\hat{N}%
_{L}(\omega),\nonumber\\
\left(  -i(\omega-\omega_{\mathrm{K}})+\frac{\kappa}{2}+\frac{\Gamma
	_{1}+\Gamma_{2}}{2}\right)  \hat{m}_{R}(\omega)+\Gamma_{1}e^{iq_{\ast}%
	|R_{2}-R_{1}|}\hat{m}_{L}(\omega)  &  =\sum_{q}ig_{q}^{\ast}e^{iqR_{2}}%
\sqrt{\kappa_{q}}G_{q}(\omega)\hat{N}_{q}(\omega)-\sqrt{\kappa}\hat{N}%
_{R}(\omega), \label{equations_noise}%
\end{align}
where $\Gamma_{1}=|g_{q_{\ast}}|^{2}/v_{q_{\ast}}$ and $\Gamma_{2}%
=|g_{-q_{\ast}}|^{2}/v_{q_{\ast}}$ are assumed constant (for the Kittel mode).
Here, $q_{\ast}$ is the positive root of $\omega_{q_{\ast}}=\omega
_{\mathrm{K}}$ as introduced in the main text.

\begin{figure}[th]
	\begin{center}
		{\includegraphics[width=9cm]{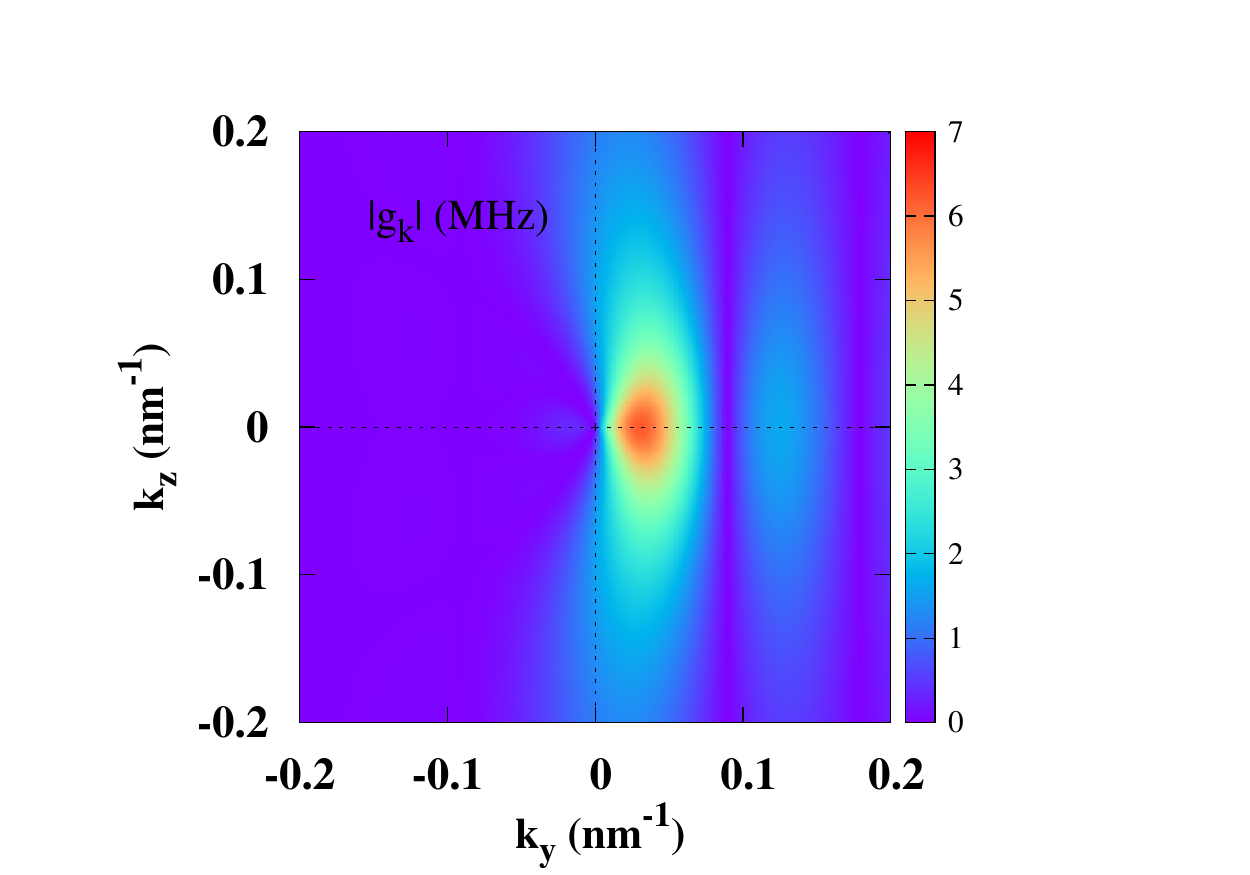}}
	\end{center}
	\caption{(Color online) Momentum dependence of the dipolar coupling strength
		$|g_{\mathbf{k}}|$ between a nanowire and magnetic film for the dimensions and
		material parameters used in the main text. \ }%
	\label{coupling}%
\end{figure}

For perfectly chiral coupling with $\Gamma_{2}=0$ the solutions of
Eqs.~(\ref{equations_noise}) read
\begin{align}
\hat{m}_{L}(\omega)  &  =\frac{\sum_{q}ig_{q}^{\ast}e^{iqR_{1}}\sqrt
	{\kappa_{q}}G_{q}(\omega)\hat{N}_{q}(\omega)-\sqrt{\kappa}\hat{N}_{L}(\omega
	)}{-i(\omega-\omega_{\mathrm{K}})+\frac{\kappa}{2}+\frac{\Gamma_{1}}{2}%
},\nonumber\\
\hat{m}_{R}(\omega)  &  =\frac{\sum_{q}ig_{q}^{\ast}e^{iqR_{2}}\sqrt
	{\kappa_{q}}G_{q}(\omega)\hat{N}_{q}(\omega)-\sqrt{\kappa}\hat{N}_{R}%
	(\omega)-\Gamma_{1}e^{q_{\ast}(R_{2}-R_{1})}\hat{m}_{L}(\omega)}%
{-i(\omega-\omega_{\mathrm{K}})+\frac{\kappa}{2}+\frac{\Gamma_{1}}{2}}.
\end{align}
With $\hat{m}_{L,R}(t)=\int e^{-i\omega t}\hat{m}_{L,R}(\omega)d\omega/(2\pi
)$, the non-equilibrium occupation of the Kittel modes becomes
\begin{align}
\rho_{L}  &  \equiv\langle\hat{m}_{L}^{\dagger}(t)\hat{m}_{L}(t)\rangle
=n_{L}+\int\frac{d\omega}{2\pi}\frac{\kappa}{(\omega-\omega_{\mathrm{K}}%
	)^{2}+(\kappa/2+\Gamma_{1}/2)^{2}}(n_{q_{\ast}}-n_{L}),\\
\rho_{R}  &  \equiv\langle\hat{m}_{R}^{\dagger}(t)\hat{m}_{R}(t)\rangle
=n_{R}+\int\frac{d\omega}{2\pi}\frac{\Gamma_{1}^{2}\kappa}{\left[
	(\omega-\omega_{\mathrm{K}})^{2}+(\kappa/2+\Gamma_{1}/2)^{2}\right]  ^{2}%
}(n_{L}-n_{q_{\ast}}),
\end{align}
where the damping in the film has been disregarded $\left(  \kappa
_{q}\rightarrow0\right)  $. In the linear regime the non-local thermal
injection of magnons into the right transducer by the left one then reads
\begin{align}
\delta\rho_{R}  &  =\left\{
\begin{array}
[c]{c}%
\mathcal{S}_{\mathrm{CSSE}}(T_{L}-T_{R})\\
0
\end{array}
\text{ when }%
\begin{array}
[c]{c}%
T_{L}>T_{R}\\
T_{L}\leq T_{R}%
\end{array}
\right.  ,\nonumber\\
\mathcal{S}_{_{\mathrm{CSSE}}}  &  =\int\frac{d\omega}{2\pi}\frac{\Gamma
	_{1}^{2}\kappa}{\left[  (\omega-\omega_{\mathrm{K}})^{2}+(\kappa/2+\Gamma
	_{1}/2)^{2}\right]  ^{2}}\left.  \frac{dn_{L}}{dT}\right\vert _{T=\left(
	T_{L}+T_{R}\right)  /2}.
\end{align}
where we defined the chiral (or dipolar) spin Seebeck coefficient
$\mathcal{S}_{_{\mathrm{CSSE}}}.$

The magnon diode effect acts a \textquotedblleft Maxwell
demon\textquotedblright\ that rectifies fluctuations in the wire temperature.
Of course, in thermal equilibrium all right and left moving magnons are
eventually connected by reflection of spin waves at the edges and absorption
and re-emission by connected heat baths. The Second Law of thermodynamics is
therefore safe, but it might be interesting to search for chirality-induced
transient effects.

\end{widetext}

\end{document}